\documentclass[aps,prc,twocolumn,groupedaddress,nofootinbib]{revtex4}

\usepackage{graphicx}
\usepackage{graphics}
\usepackage{verbatim}
\usepackage{color}  
\usepackage{soul}
\usepackage{amsfonts,amsmath,amssymb}

\usepackage{dcolumn}
\newcolumntype{d}[1]{D{.}{.}{#1}}

\begin{document}

\title{Benchmark of many-body approaches for magnetic dipole transition strength}

\author{M.~Frosini}
\address{CEA, DES, IRESNE, DER, SPRC, LEPh, 13115 Saint-Paul-lez-Durance, France}
\author{W.~Ryssens}
\address{Institut d'Astronomie et d'Astrophysique, Universit\'e Libre de Bruxelles, CP-226, 1050 Brussels, Belgium}
\author{K.~Sieja}

\address{Universit\'e de Strasbourg, IPHC, 23 rue du Loess 67037 Strasbourg, France\\
CNRS, UMR7178, 67037 Strasbourg, France}
\email{kamila.sieja@iphc.cnrs.fr}
\date{\today}

\begin{abstract}
\begin{description}
\item[Background:] The low-energy enhancement observed recently in the deexcitation $\gamma$-ray strength functions, suggested to arise due to the magnetic dipole ($M1$) radiation, motivates theoretical efforts to improve the description of $M1$ strength in available nuclear structure models. Reliable theoretical predictions of nuclear dipole excitations are of interest for different nuclear applications and in particular for nuclear astrophysics, where the calculations of radiative capture cross sections often resort to theoretical $\gamma$ strength functions.
\item[Purpose:] We aim to benchmark many-body methods in their description of the $M1$ $\gamma$ strength functions, with a special emphasis on the description of the low-energy effects observed in the deexcitation strength.   
\item[Methods:] We investigate the zero-temperature and finite-temperature (FT) magnetic dipole strength functions computed within the quasiparticle random-phase approximation (QRPA) and compare them to those from exact diagonalizations of the same Hamiltonian in restricted orbital spaces. The study is carried out for a sample of 25 spherical and deformed nuclei, with masses ranging from $A=26$ to $A=136$, which can be described by diagonalization of the respective effective Hamiltonian in three different valence spaces. 
\item[Results:] A reasonable agreement is found 
for the total photoabsorption strengths while the QRPA distributions are shown to be
systematically shifted down in energy with respect to exact results. Photoemission strengths obtained within the FT-QRPA formalism
appear insufficient to explain the low-energy enhancement of the $M1$ strength functions
present within the exact diagonalization approach.
\item[Conclusions:] The problems encountered in the zero- and finite-temperature QRPA 
calculations are ascribed to the lack of correlations in the nuclear ground state and to the truncation of the many-body space. In particular, the latter prevents obtaining the sufficiently high level density to produce the low-energy enhancement of the $M1$ strength function, making the (FT-)QRPA approach unsuitable for predictions of such effects across the nuclear chart.

\end{description}
\end{abstract}
%

\maketitle

\section{Introduction}

Radiative neutron capture plays a crucial role in many applications of nuclear
physics, from reactor design to astrophysical simulations. Experiments cannot 
realistically obtain the cross section of this process in 
all relevant conditions or even for all relevant nuclei; in the case of r-process
nucleosynthesis thousands of extremely exotic nuclei far beyond the reach of
current accelerators are involved~\cite{Arnould2020}. Reaction theory can 
provide the missing information through the statistical Hauser-Feshbach model, 
but the resulting cross section depends strongly on the structure of the compound
nucleus formed, and in particular on its probability to deexcite through the 
emission of a $\gamma$-ray~\cite{Hauser-Feshbach}. This probability is characterized by the
$\gamma$-ray strength function ({$\gamma$SF}) of the nucleus, a quantity that 
should ideally be predicted reliably across the entire nuclear chart by 
nuclear structure models.

Although radiation of all multipolarities can contribute to
the strength functions, the main contributions to neutron capture cross sections 
are due to dipole radiation: both the giant electric dipole resonance at high 
energy and the magnetic spin-flip resonance at low energy.
Traditionally, dipole strength functions were modeled in the fully phenomenological 
Lorentzian approximations \cite{SLO, RIPL-3}. However, important deviations 
from statistical behavior were evidenced at low $\gamma$-ray energies, in 
particular the so-called low-energy enhancement (LEE) of the dipole 
strength~\cite{oslo-website, Larsen-Fe56, enh-Mo, end-Sc43, Larsen-Ti44, Larsen2010}. 
Such an enhancement, if present in neutron-rich nuclei, could increase 
the neutron-capture cross sections up to a 100 times~\cite{Larsen2010}, depending
on whether the electric or magnetic dipole strength is largest. 

Significant theoretical effort was devoted in recent years to explain the LEE.
The authors of Ref.~\cite{Schwengner-Mo} achieved an initial breakthrough: for the first time, 
the deexcitation strength function was obtained from interacting shell-model (SM) 
calculations, pointing to the magnetic-dipole character of the LEE. Further 
SM calculations of both the $M1$ and $E1$ dipole strength functions of $^{44}$Sc 
confirmed the magnetic-dipole nature \cite{Sieja-PRL} of the LEE and predicted a flat behavior
of the $E1$ strength functions towards zero $\gamma$-ray energy. Further SM studies
adressed the LEE more systematically and concluded that the enhancement is produced
by the low-energy $\gamma$ rays coming from the quasi-continuum of nuclear 
states~\cite{Schwengner-Fe, Sieja-EPJA, Brown-V, LEMAR, Karampagia2018, Mitbo-LEE}.

Other studies of the LEE so far have been based on the (quasiparticle) 
random phase approximation or (Q)RPA but typically generalize the traditional 
QRPA formalism by including for instance the effect of finite temperature. 
Two early studies were limited to the electric dipole strength function~\cite{elena-Mo, Wibowo2019},
while a more recent study addressed both dipole modes in $^{56}$Fe~\cite{Beaujeault2022}. 
The combined deexcitation strength of $E1$ and $M1$ obtained by the authors of 
the latter appeared insufficient to describe the low-energy data from Oslo 
experiments~\cite{oslo-website}, while shell-model calculations achieved
good agreement with only the magnetic decay strength~\cite{brown-Fe56}.

The interacting shell model with highly-tuned empirical Hamiltonians is known to provide
precise results for spectroscopy and electromagnetic transitions. Unfortunately due to computational complexity
its applications are still restricted to particular regions of nuclei. The 
necessity of deriving a suitable effective interaction for each model space adds
to the already complicated task of the complete diagonalization of the many-body
Hamiltonian, making it impossible to achieve systematic studies that span the 
nuclear chart. The (quasiparticle) random phase approximation or (Q)RPA approach 
provides an interesting alternative: this method scales polynomially with particle
number thanks to a truncation of the many-body space to either two quasiparticle 
excitations (QRPA) or particle-hole excitations (RPA) of a mean-field reference state. Because of this favorable scaling, 
(Q)RPA and its extensions~\cite{Ring80} have been widely used in many different 
contexts~\cite{Paar2007}, from \textit{ab initio} interactions to systematic 
studies across the nuclear chart with energy density functionals
~\cite{Terasaki2010,Martini_2016,Krusiv2021}.

However, the truncation of the many-body space necessarily misses physical effects
present in a complete SM calculation such as the LEE. More advanced many-body 
approaches aim to decrease such errors by including higher order excitations 
(two-particle two-hole (2p-2h), 3p-3h), which enhances the fragmentation of the 
spectrum while shifting the centroid of the resonance~\cite{Gambacurta2012,Gambacurta2015,Knapp_2023,trippel2016}. 
However, such calculations are generally very demanding and are impractical 
for global application. A more pragmatic approach to provide a complete set of 
dipole $\gamma$-ray strength functions derived from QRPA calculations was 
developed in Refs. \cite{Goriely2002, Martini2016, Goriely2016}
by adding further empirical corrections to account for the missing correlations
and to reproduce the available data. Ref.~\cite{Goriely2018} in particular 
provided a complete set of dipole strength functions that include low-energy 
structure effects by phenomenological corrections inspired by SM calculations. 
Such a treatment appeared successful but surely is far from being satisfying if 
one aims at a fully coherent microscopic description of strength functions across
the nuclear chart: a limited number of available shell-model results does 
not guarantee the universality of the observed low-energy effects which, if 
applied globally through a phenomenological recipe, may introduce unrealistic 
behaviors of the neutron-capture cross sections for exotic nuclei.

Although SM calculations cannot cover sizable portions of the nuclear chart, 
their exact nature is the ideal benchmark of approximate methods that scale
more gently such as the QRPA \cite{Goriely2018,sieja2020directcapture}. Our goal 
here is to understand in more detail the deficiencies of the QRPA approach: we 
study the differences between strength functions obtained with exact 
diagonalization and QRPA calculations in identical model spaces and employing 
identical SM Hamiltonians. Ref.~\cite{Stetcu2003} constitutes a previous benchmark along this line: 
the authors studied a number of transition operators (Gamow-Teller, spin-flip and quadrupole)
but limited themselves to a few nuclei in the $sd$ and $pf$ shells and 
RPA calculations that did not include the effects of pairing. 
Here we concentrate on the magnetic dipole operator, include the effect of
pairing by utilising the QRPA and cover a wider range of 
nuclei, with masses from $A=20$ to $136$, in three distinct model spaces.
Standard QRPA by default only provides \emph{photoabsorption} strength functions; 
we extend the benchmark to finite-temperature QRPA 
(FT-QRPA), which is arguably the simplest possible extension of QRPA that 
offers access to the \emph{photoemission} strength function and that has been used to 
to study the LEE~\cite{elena-Mo}.

This paper is organized as follows: we remind the reader of the basics of both 
theoretical approaches in Section \ref{sec-theory}. We present the results for 
the nuclear ground states and photoabsorption strength in Section \ref{sec-zero} 
and discuss the origins of the discrepancies between the many-body methods. 
In Sec. \ref{sec-lee} we discuss the description of the photoemission strength 
in SM and FT-QRPA, including the temperature behavior of the computed 
strength functions and with an emphasis on the LEE. Finally, Section \ref{sec-pro} 
concerns our conclusions and perspective for future developments aimed 
at the systematic microscopic description of magnetic dipole strength functions. 

\section{Theory framework \label{sec-theory}}
In order to compare the results of different theoretical approaches
we will discuss the sum rules, centroids and widths of strength distributions
following the standard definitions~\cite{Ring80, Stetcu2003}.
Denoting the ground state and all the excited states by \(|0\rangle\) and 
\(|\nu\rangle\) respectively, the total strength 
\begin{equation}
S_0=\sum_\nu|\langle\nu|\hat O|0\rangle|^2 \, ,
\end{equation}
is the non-energy-weighted sum rule associated with a transition
operator $\hat O$. The centroid and width of this strength function are then
\begin{eqnarray}
\bar S=\frac{S_1}{S_0}, \qquad \Delta S=\sqrt{\frac{S_2}{S_0}-\bar S^2} \, ,
\end{eqnarray}
where
\begin{equation}
S_k=\sum_\nu(E_\nu-E_0)^k |\langle\nu|\hat O|0\rangle|^2 
\end{equation}
is the (energy-weighted) sum rule of the order $k$. We focus here on the magnetic dipole operator:
\begin{equation}
\hat O(M1)=\sqrt{\frac{3}{4\pi}}\sum_k \left[g_l(k){\hat l}(k)+g_s(k){\hat s}(k)\right]\mu_N \, ,
\label{eq:M1operator}
\end{equation}
where $\hat l$ and $\hat s$ are the orbital and spin angular momentum 
operators and the sum runs over all individual nucleons. The orbital and spin gyromagnetic factors
are given by $g_l$=1 and $g_s$=5.586 for protons and $g_l$=0 and $g_s$=-3.826
for neutrons. We employ these bare values for the orbital angular momentum 
but multiply the spin factors by 0.75 as is customary for calculations limited
to a valence space, see e.g. Ref.~\cite{RMP} and references therein. 

The reduced transition probability from an initial state \(|i\rangle\) to a final state \(|f\rangle\) is calculated as 
\begin{equation}
B_{fi}=\frac{1}{2J_i+1}\langle f||\hat O||i\rangle^2\,. 
\end{equation}

The $B(M1)$ distributions are convoluted with Lorentzians of an arbitrary width $2\gamma=1$MeV
and converted into photo-strength function (in units of MeV$^{-3}$) according to the formula \cite{Loens}
\begin{equation}
f_{M1}=16\pi/27(\hbar c )^3\sum_f B_{fi}(M1)\frac{1}{\pi}\frac{\gamma}{(E-\Delta(E_{fi})^2+(\gamma)^2}
\end{equation}
which leads to the continuous strengths presented in Figures \ref{fig-xe108pn}-\ref{fig-qrpa-abs} and \ref{fig-decay}.

\subsection{Model-space and Hamiltonian}

Calculations are carried out using model spaces with well-established empirical interactions that are capable of describing (with a full diagonalization) the low-energy levels of nuclei
within the major shell with an accuracy of around 200 keV:
USDb \cite{USDB} for the $1s0d$ shell, LNPS \cite{Lenzi2010} for the 
$1p0f$ shell and GCN5082 interaction \cite{Gniady, xenon} in the $0g_{7/2}1d2s_{1/2}0h_{11/2}$ shell.
As the full model-space diagonalizations become quickly difficult/impossible
with the number of valence particles in the $0g_{7/2}1d2s_{1/2}0h_{11/2}$ model space, 
only a few nuclei close to the $N=Z$ line and close to  the $N=82$
shell closure are considered (roughly the same nuclei for which the radiative decay was previously
studied within the shell-model framework in \cite{Sieja2018}). 
The same quenching factor of 0.75 is applied on the spin part of the magnetic operator in all model spaces,
even though more sophisticated prescriptions exist for a better agreement with experiment
(see e.g. \cite{PhysRevC.65.024316}). However, the choice of effective operators does not
play any role in this study aiming only in comparison of theoretical models.

\subsection{Exact diagonalization}
The reference results in this work are obtained in the shell-model framework, i.e. by diagonalization of the Hamiltonian in the basis of many-body states that can be constructed by placing $n$ nucleons in the valence-space orbitals. We will dub those results hereafter as exact or shell-model results. 
Distributions of $B(M1)$ strengths in the shell model are computed using Lanczos strength functions method
which permits to get the strength per energy interval in an efficient way \cite{RMP}. 
We remind that the choice of the starting vector, called pivot, used in the Lanczos diagonalization procedure is arbitrary.
Given a transition operator $\hat O$ one can define a pivot of the form $\hat O |\Psi_i\rangle$,
where $|\Psi_i\rangle$ can be chosen any shell-model state, and carry on Lanczos diagonalization.
The unitary matrix $U_{ij}$ that diagonalizes the Hamiltonian after N Lanczos iterations contains then   
in its first row the amplitude of the pivot in the $j$th eigenstate. Thus $U_{1j}^2$ as function of 
eigenergies $E_j$ defines the strength function of the pivot state.
Note that to obtain the total strength $S_0$ for the ground state to be compared to QRPA
only diagonalization of the $0^+$ state has to be carried out in even-even nuclei, 
as the sum rule is the norm of the pivot state obtained 
by acting with the transition operator on the initial state.
The remaining moments of the distribution presented in Tables are extracted from the peaked-fence
distributions obtained with the Lanczos strength function method with 100 iterations.
These calculations are done using the m-scheme shell-model code ANTOINE \cite{ANTOINE, RMP}.
In addition to photoabsorption strength, the decay strength functions are also computed  
employing the Bartholomew definition \cite{Bartholomew}, following Refs. \cite{Sieja-PRL, Mitbo-LEE}:
\begin{equation}
f_{M1}(E_{\gamma}, E_i, J_i, \pi)=16\pi/9(\hbar c )^3 \langle B(M1)\rangle \rho(E_i,J_i, \pi),
\label{eq-fm1}
\end{equation}
where $\rho_i(E_i,J_i,\pi)$ is the partial level density determined at a given initial excitation energy $E_i$ and $\langle B(M1)\rangle$ averaged reduced transition probability per energy bin. 
As such a calculation requires computation of hundreds of converged excited states, 
the $j$-coupled code NATHAN \cite{RMP} is employed to achieve this task and avoid numerical problems appearing
in the $m$-scheme where large number of Lanczos iterations is necessary \cite{RMP}. 
The details about energy and spin cutoffs of these calculations are given in Sec. \ref{sec-lee} for each of considered nuclei.

\subsection{QRPA at zero and finite temperature}
\label{sec:qrpa}
Finite temperature QRPA (FT-QRPA) builds on top of finite temperature Hartree-Fock Bogoliubov (FT-HFB) 
calculations where mean-field pure product state is replaced by a statistical mixture of Bogoliubov states 
characterized by the one-body density operator. 
FT-HFB state is obtained by minimizing the grand-canonical potential defined in 
Ref.~\cite{HF-SHELL}. In particular, the FT-HFB state \(|\Phi(T)\rangle\) is fully determined by a Bogoliubov transformation defining the set of quasiparticle operators \(\{\beta,\beta^\dag\}\) from single-particle operators \(\{c,c^\dag\}\)
\begin{equation}
    \begin{pmatrix}
        \beta\\\beta^\dag
    \end{pmatrix}
    =\begin{pmatrix}
        U&V^*\\V&U^*
    \end{pmatrix}^\dag
    \begin{pmatrix}
        c\\c^\dag
    \end{pmatrix} \, ,
\end{equation}
and occupation numbers \(f\) such that the generalized density matrix \(\mathcal R_0\) is diagonal in the quasiparticle basis, i.e. it reads
\begin{equation}
    \mathcal R_0 = \begin{pmatrix}
        f&0\\0&1-f
    \end{pmatrix} \, .
\end{equation}
Occupation number \(f_\mu\) is related to quasiparticle energies \(E_\mu\) by
\begin{equation}
    f_\mu\equiv \frac{1}{1+e^{\frac{E_\mu}{k_BT}}}.
\end{equation}

FT-QRPA approximates the ground and excited states by performing elementary 2 quasiparticle excitations around \(|\Phi(T)\rangle\). 
FT-QRPA thermal excited states \(|\mu\rangle\equiv \Gamma_\mu^\dag|\Phi(T)\rangle\) are parametrized by finite temperature amplitudes \(X^\mu, Y^\mu, P^\mu, Q^\mu\) as
\begin{equation}
    \Gamma_\mu^\dag \equiv \frac12\sum_{ij} \left[ 
    P^\mu_{ij}\beta^\dag_i\beta_j
    +X^\mu_{ij}\beta^\dag_i\beta^\dag_j
    -Y^\mu_{ij}\beta_j\beta_i 
    -Q^\mu_{ij}\beta_j \beta^\dag_i
    \right].
\end{equation}
Expressions
 for the amplitudes are obtained as solution of an eigenvalue equation derived equivalently from linearized TD-FT-HFB equations~\cite{SOM83,HF-SHELL2} 
 or from linearized equation of motion~\cite{Rowe70,SOM83}.

FT-QRPA contains two different approximations that will make it deviate from the exact diagonalization:
\begin{itemize}
    \item Limitation to the space of 2 quasiparticle states, preventing in particular explicit p-n correlations and restoration of broken symmetries,
    \item Quasi-boson approximation~\cite{Rowe70}, emerging from replacing the correlated FT-ground state by \(|\Phi(T)\rangle\) in the evaluation of nested commutators and causing a violation of Pauli exclusion principle.
\end{itemize}


Given a one-body transition operator \(F\)\footnote{In the case of M1 transition, \(F\equiv M^{1\mu}\).}, the susceptibility
 \(\chi_F\) is defined as
\begin{equation}
    \chi_F(\omega)\equiv \langle \Phi(T) | [\Gamma(\omega), F] | \Phi(T) \rangle,
\end{equation}
where \(\Gamma(\omega)\equiv \sum_\mu \frac{\Gamma_\mu}{\omega-\Omega_\mu} \) and \(\Omega_\mu\) are the FT-QRPA poles. Following this definition
the FT-QRPA excitation strength function is expressed as
\begin{equation}\label{eq:prefact}
    S_F(\omega) \equiv - \frac1{\pi(1-e^{-\beta\omega})} \mathrm{Im} \chi_F(\omega).
\end{equation}
This strength contains both an absorption (\(\omega>0\)) and a deexcitation part (\(\omega<0\)) that will be considered when studying the LEE.


Zero-temperature QRPA is naturally obtained as a limiting case of FT-QRPA, where all \(f\) identically vanish along with \(P^\mu\text{ and }Q^\mu\) amplitudes. This also means that the dimensionality of FT-QRPA is twice larger as zero-T QRPA. In practice, only states close to the Fermi energy are unblocked at low temperature, which tends to enrich the QRPA strength at low energies via the apparition of low-lying poles.
The zero-T limit of Eq.~\eqref{eq:prefact} only retains transitions from ground to excited states while the thermal prefactor becomes a step function by making the deexcitation part vanish identically.
In the rest of this work, zero-T QRPA is referred to as QRPA.

In the present work, we employ the recent numerical implementation of the Finite Amplitude Method (FAM) for solving FT-QRPA equations. FT-QRPA-FAM replaces the intensive calculation 
and diagonalization of the FT-QRPA matrix by a set of non-linear equations of similar 
dimension to that of the static Hartree-Fock-Bogoliubov mean-field approach it builds upon. 
The QRPA-FAM, first proposed in \cite{Nakatsukasa2007, Nakatsukasa2011}, 
has proven to be a very efficient tool to obtain electric \cite{Inakura2009, Stoitsov2011} and charge-exchange 
\cite{Ney2020, Mustonen2014} strength functions, 
as well as to determine collective inertia \cite{Washiyama2021}, quasiparticle-vibration coupling \cite{Litvinova2021}, 
discrete eigenmodes \cite{Hinohara2013} and sum rules \cite{Hinohara2015}.
In a recent work \cite{Beaujeault2022}
a QRPA-FAM implementation to compute zero- and finite-temperature 
strength functions using {\it ab-initio} interaction was presented.
Here we use the same numerical implementation
to study dipole strength functions but with shell-model Hamiltonians.
In particular, the present QRPA-FAM implementation permits to go beyond the axially-deformed 
approach, which we explore in Sec. \ref{sec-triax}. 

Axially symmetric QRPA-FAM calculations have been benchmarked with numerical implementation of the matricial FT-QRPA
formalism presented in \cite{HF-SHELL2} to the HF-SHELL
code published in Ref. \cite{HF-SHELL}.
Both implementations match exactly. The results at (finite T) zero T will be indifferently referred to as (FT-)QRPA calculations in the following. Only even-even nuclei are computed in this work with FT-QRPA, an extension to odd systems is envisioned.



\section{Ground states: absorption strength}
\label{sec-zero}

To test the QRPA approaches in the description of magnetic dipole strengths
we use a set of 25 spherical and deformed nuclei which can be described
in the $1s0d$, $1p0f$ and $0g_{7/2}1d2s_{1/2}0h_{11/2}$ spaces within the shell-model approach
by exact diagonalization in the full model space. 
The same orbital spaces with their respective effective Hamiltonians 
are then used to perform calculations of transition strengths
within the QRPA framework. 

\subsection{Mean-field solutions}

The starting point of all (FT-)~QRPA calculations is a mean-field state: 
we construct either (i) the Hartree-Fock-Bogoliubov (HFB) state that minimizes the total 
energy at zero temperature or (ii) the statistical mixture of HFB 
states that minimizes the free energy at finite temperature~\cite{HF-SHELL}.
These configurations are constructed from single-particle states of definite
proton or neutron nature expanded in the basis of the valence space orbitals, 
i.e. we do not allow for isospin mixing between protons and neutrons.
We allow for the spontaneous breaking of rotational and particle number symmetry, 
but restrict ourselves to axial symmetry except when explicitly mentioned. 
We summarize the results of our zero-temperature HFB calculations in Table \ref{tab-1}: 
we list ground state energies from the exact diagonalization ($E_{\rm SM}$) and 
the difference with respect to the HFB states ($\Delta E$). To gauge the 
degree of symmetry breaking present in our mean-field configurations, we 
also include the quadrupole deformation $\beta_{20}$ and whether or not the
neutrons or protons form a pair condensate at the mean-field level
\footnote{The values of $\beta_{20}$ we list should not be compared to 
         values extracted from experimental data. The values we present reflect
         only the deformation of the valence nucleons, while intrinsic 
         quadrupole deformation is a collective phenomenon that naturally includes 
         contributions from all nucleons. We did not include a rescaling 
         of the quadrupole operator to account for this effect here for simplicity.}.

As can be seen from the Table, the results for the ground state energy fall into
two groups: of light nuclei ($sd$-shell and $pf$-shell) where the disagreement 
is large and of the heavier nuclei ($gdsh$-shell) where difference between 
exact and HFB energies is smaller, especially when compared to the total binding
energy. In most cases, nuclei computed with valence space 
interactions do not break simultaneously particle number and rotational symmetry. 
If the former symmetry is not spontaneously broken, the HFB formalism reduces
in practice to the Hartree-Fock formalism while QRPA reduces to RPA~\footnote{
When pairing vanishes, QRPA reduces to the combination of 
RPA and PP-RPA: the first dealing with 1p-1h excitations and the second 
with 2p and 2h excitations~\cite{RingSchuck}. We ignore this subtlety here, since the matrix 
element of the $M1$ operator between two Slater determinants with different particle number
vanishes and hence does not contribute to strength functions. In general
however, the zero-pairing limit of HFB and QRPA approaches should be treated
with care~\cite{duguet2020,duguet2020a}.}. In what follows, we will use QRPA
indiscriminately in all cases except when explicitly mentioned.
\begin{table}
\caption{Ground state properties of the nuclei considered in this study, 
organised by the corresponding valence space. We list minus the binding energy 
$E_{\rm SM}$ obtained with exact diagonalization (in MeV) and the energy difference 
with respect to (zero-temperature) HFB calculations $\Delta E \equiv E_{\rm SM} - E_{\rm HFB}$ (in 
MeV) as well as the quadrupole deformation $\beta_{20}$. The last two columns indicate whether or not 
pairing correlations are present in the HFB solution for each nucleon species (Y=yes, N=no).}
\label{tab-1}
\begin{center}
\begin{tabular}{cccccc}
\hline
Nucleus & $E_{\rm SM}$  & $\Delta E$ & $\beta_{20}$ & $E_{\rm p}$ & $E_{\rm n}$\\
\hline
$^{20}$Ne  &-40.47 &-4.07&0.30   & N & N \\  
$^{24}$Ne  &-71.72 &-5.32 &0.18  & N & Y \\
$^{24}$Mg  &-87.10 &-6.34&0.27   & N & N  \\ 
$^{28}$Mg  &-120.49 &-4.87&0.18  & N & N  \\
$^{28}$Si  &-135.86 &-5.84&-0.24 & N & N \\  
$^{32}$Si  &-170.52 &-4.18&-0.13 & N & N \\   
$^{32}$S   &-182.44 &-6.05&0.00    & N & N\\   
$^{36}$Ar  &-230.27 &-3.61&-0.11 & N & N\\ 
\hline
$^{44}$Ti   &-46.88&-3.65 &0.12  & N & N\\
$^{50}$Ti   &-108.68&-4.34&0.00    & Y & N\\
$^{48}$Cr   &-98.72&-4.67&0.16    & N & N \\
$^{52}$Cr   &-142.88&-4.08&0.05   & Y & N\\
$^{52}$Fe   &-151.64&-6.51 &0.12 & N & N \\
$^{56}$Fe   &-195.40&-7.27 &0.12 & N & Y \\
$^{56}$Ni   &-205.92&-6.36 &0.00   & N & N \\
$^{60}$Ni   &-248.04&-6.64 &0.00   & N & Y \\
$^{64}$Zn   &-303.02&-6.81&-0.15 & Y & N\\
$^{64}$Ge   &-310.84&-8.56&-0.15 & N & N \\
\hline    
$^{104}$Te &-50.26&-2.23  &0.05  & N & N \\
$^{108}$Te &-98.05&-2.68  &0.07  & N & Y \\
$^{108}$Xe &-102.52&-4.09  &0.08  & Y & Y \\
$^{128}$Te &-282.14&-2.53  &0.03 & Y & Y \\
$^{132}$Te &-309.51&-1.25  &0.00   & Y & Y \\
$^{134}$Xe &-353.22&-2.08 &-0.02 & Y & Y \\
$^{136}$Ba &-396.02&-2.61 &-0.03 & Y & Y \\
\hline      
\end{tabular}
%
\end{center}
\end{table}

\subsection{$M1$ dipole response \label{sec-M1}} 
\begin{figure*}[t]
\includegraphics[width=0.3\textwidth]{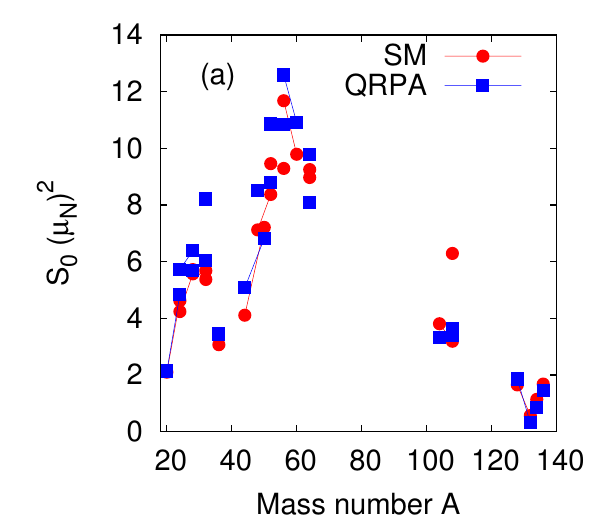}
\includegraphics[width=0.3\textwidth]{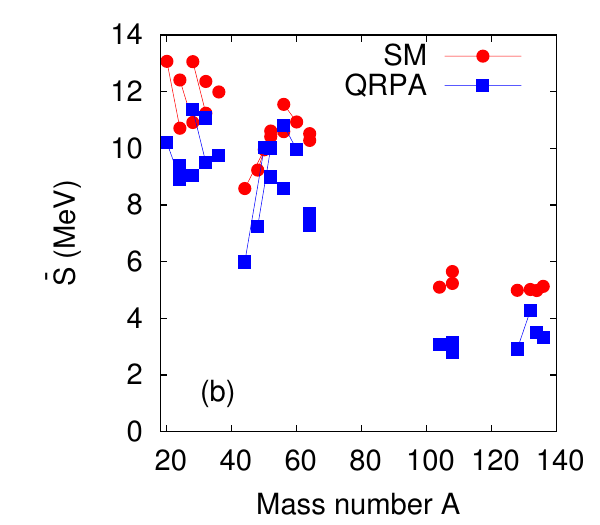}
\includegraphics[width=0.3\textwidth]{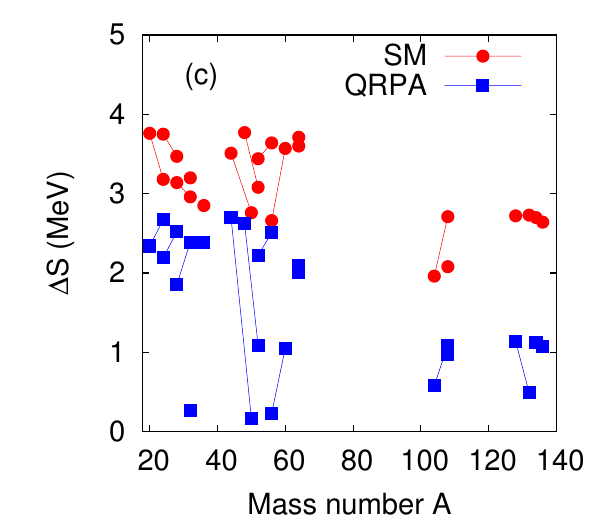}
\caption{Total strengths (a), centroids (b) and widths (c)
for magnetic dipole operator as obtained in SM (red) and QRPA (blue) 
approaches for nuclei listed in Table \ref{tab-1}.
The lines connect the same $Z$-numbers.}
 \label{fig-sr}
\end{figure*}

The characteristics of the $B(M1)$ distributions
for all computed nuclei are plotted in Fig. \ref{fig-sr} 
comparing shell model to the axially-deformed QRPA calculations.
The general trends are easy to note and independent on the model space/ Hamiltonian employed.
The total strengths (panel (a) of the Figure) exhibit the same tendencies in both approaches and agree within $20\%$
for the majority of nuclei. 
One of the largest discrepancies, well visible in Fig. \ref{fig-sr} around
$A=100$, concerns the $N=Z$ $^{108}$Xe nucleus, which  
was predicted in a previous shell-model study with the same interaction to be triaxially deformed
with $\beta=0.16$ and $\gamma=24^{\circ}$ \cite{xenon}. 
The QRPA is missing nearly twice the strength predicted in the shell model in this case.
Interestingly, in the other triaxial nucleus, $^{24}$Mg, the QRPA sum rule overshoots the shell-model value
by $35\%$. Thus the triaxiality itself is not the reason behind the missing strength 
observed in $^{108}$Xe. The symmetry-unrestricted calculations verifying the
actual impact of non-axiality on the strength distributions
in these two nuclei are presented in Sec. \ref{sec-triax}.
   
While the total strength seems reasonably reproduced by QRPA with a few exceptions,
the centroids are always shifted to lower energies than the SM ones and the QRPA distributions
are less spread. This appears to be a common feature of the QRPA method for all transition operators
as one can conclude comparing our results and those of Ref. \cite{Stetcu2003}. 
One can also note that the width of the strength distributions are worse reproduced in spherical nuclei.
While the small fragmentation of QRPA strengths can be attributed to the lack of higher-order particle-hole
correlations, the shift of the centroid is more troublesome and additionally does not seem correlated with
quadrupole deformation. The authors of Ref. \cite{Stetcu2003} 
suspected inclusion of pairing within HFB+QRPA would improve the situation: as can be taken from
our results, pairing correlations are not sufficient to cure the general shift of the QRPA distributions
to lower energies. One can note from Fig. \ref{fig-sr} that the behavior of centroids and widths
is also the same in all studied regions, 
while only in the heaviest nuclei truly paired HFB mean-field solutions are obtained.
The influence of pairing is however addressed in more detail in Sec. \ref{sec-pair}
where the solutions with/without pairing correlations in selected nuclei are discussed.

The second hypothesis addressed in Ref. \cite{Stetcu2003} was that the missing low-energy strength 
is due to the incomplete restoration of the symmetries in the RPA. For quadrupole strength, that is naturally impacted by rotational properties of the nucleus, the argument is indeed very well plausible and confirmed by other studies \cite{porrothesis}. However, we suspect this does not hold for $M1$ transitions that are not expected to be of rotational character. 

A possible explanation of the regular shift of the centroid towards lower energies 
can be traced back to the correlations missing in the mean-field treatments of ground and excited states.
In Ref. \cite{Beaujeault2022} the effect of introducing correlations beyond HF+RPA 
on $E1$ photoabsorption cross sections was discussed with {\it ab-initio} interactions.
It was shown that introducing correlations to the ground state via coupled cluster RPA or in-medium
RPA leads to shifting of the whole strength in $^{16}$O by 5-6MeV providing a desired effect in view of our results.
However, adding correlations also to the excited
states through the second RPA method pushes the whole strength down by the same amount, resulting
in photoabsorption cross section closer to the initial HF+RPA result.  
To get more insight into the correlations present in different approaches, 
we have computed occupations of the spherical orbits 
resulting the HF(B) calculations of the ground states for selected nuclei where the disagreement with the shell-model diagonalization is particularily large. Those are compared
to occupations from exact diagonalization in Table \ref{tab-occup}. 
Since we consider $N=Z$ nuclei
and the interactions are isospin conserving, the proton and neutron occupations are equal thus only one of them 
is reported in the Table. We have also truncated 
shell-model calculations in order to get similar occupations as in HF(B) -
those are indicated in the Table as SM$^{mod}$. 
Further, we recomputed the $M1$ sum rules on such modified ground- 
state wave functions and used them as pivots in the Lanczos strength function method. 
The values of total strengths, centroids and widths obtained
with such modified shell-model wave functions
are given in Table \ref{tab-mod} together with QRPA and full-space diagonalization results. 

\begin{figure}
\includegraphics[width=0.4\textwidth]{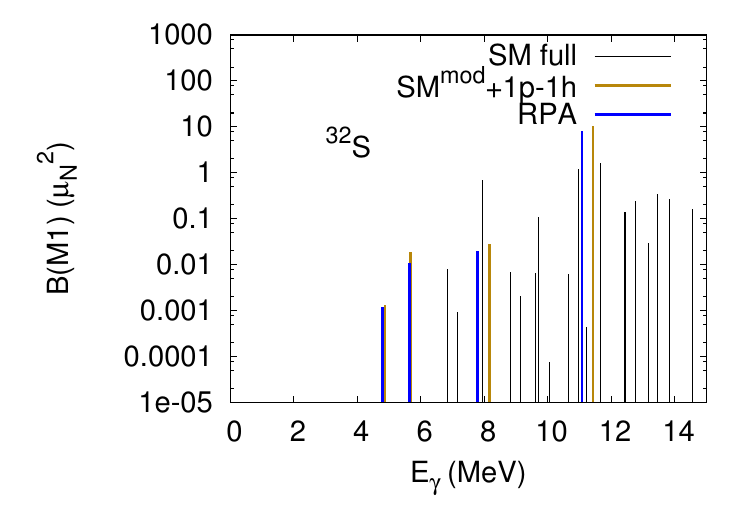}
\includegraphics[width=0.4\textwidth]{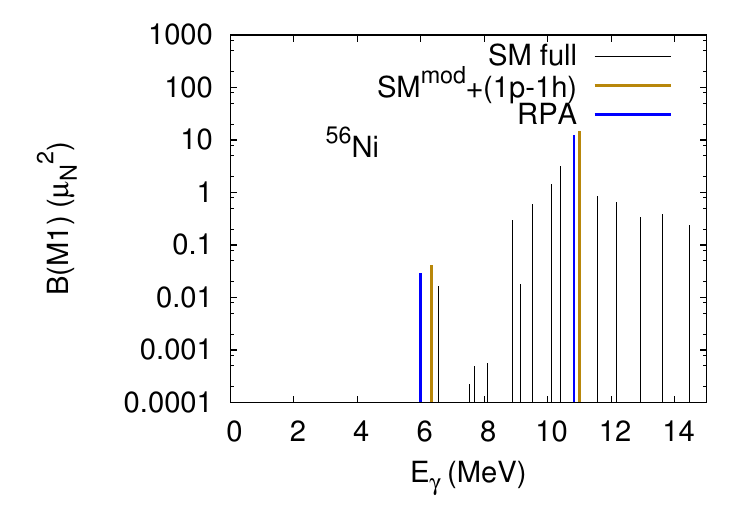}
\caption{$M1$ strength in RPA, SM and modified SM calculations in $^{32}$S and $^{56}$Ni. See text for details.\label{fig-trunc}}
\end{figure}

Taking first as examples spherical nuclei 
$^{32}$S and $^{56}$Ni, the Hartree-Fock wave functions are simply the lowest-filling configurations
without correlations which are present in the shell-model solutions as seen from the Table and the QRPA reduces to RPA in this case.
We have thus truncated the SM configuration space to force the $0^+$ states to be 0p-0h configurations with respect to the reference Slater determinant 
and then allowed for maximally 1p-1h excitations to the remaining orbits for both protons and neutrons to describe excited $1^+$ states.
A comparison of the shell-model $M1$ strength obtained in full and truncated model space is shown
in Fig. \ref{fig-trunc}, together with the RPA results, while the values characterizing these distributions are given in Tab. \ref{tab-mod}.
As can be seen, in $^{32}$S the $M1$ strength in RPA calculations is concentrated in a single peak 
at 11.1MeV with 3 other states predicted by the theory that carry very little strength. The diagonalization also gives 4 states at similar energies, with one major peak at 11.4MeV. As one can see in the Table, the diagonalization predicts however larger total strength but the centroid and width are very close to the RPA values. Similarly, in $^{56}$Ni the RPA gives two peaks, the one at 10.82MeV carrying 99\% of the $M1$ strength, in a good agreement with the restricted-space diagonalization, though the total strength is larger in the latter. 
Since the ground state correlations and particle-hole content of excited states is now the same, 
the remaining difference between SM$^{mod}$+(1p-1h) and RPA 
most likely comes from the quasi-boson approximation \cite{Ring80}. 

Contrary to the spherical nuclei $^{32}$S and $^{56}$Ni,  
in $^{104}$Te the Hartree-Fock solution is much closer to that of the diagonalization
though the $0h_{11/2}$ orbital remains empty in HF while 0.1 particle is occupying this orbital in the SM.
The diagonalization performed preventing
the particles to be promoted to the $0h_{11/2}$ orbital gives very similar occupations to the HF solution, see Tab. \ref{tab-occup},
one can thus suppose the ground-state correlations are equally taken into account in the RPA and the SM$^{mod}$. Performing strength function calculations without any further restriction on the structure of excited states one recovers the total RPA strength in $^{104}$Te, see Table \ref{tab-mod}.
Still, the centroid and width of the distribution with a modified ground state are in-between the RPA
and full SM values meaning the approximations made in the RPA to describe 
excited phonon states are insufficient. 
Adding more nucleons in $^{108}$Xe non-trivial pairing solutions are obtained in the ground state resulting in occupation of the 
$0h_{11/2}$ orbital of 0.08 particle versus 0.3 particle in the exact wave function. 
Repeating the exercise for $^{108}$Xe to get similar orbital occupancy in shell model and HFB ground states,
the total strength from the exact solution goes lower without 
populating the $0h_{11/2}$ and thus gets closer to the QRPA value.  
The conclusions remain however the same as in $^{104}$Te, in spite of pairing interactions 
additionally taken into account this time. 
These calculations put in evidence a crucial role of inclusion of correlations in the ground state
and in the excited states simultaneously to reproduce the centroid and width of the distribution. It is additionally shown (for the first 2 studied cases) that the quasi-boson approximation introduces an additional inaccuracy to the calculation of the QRPA strength.

\begin{table}
\caption {Occupation of spherical orbits resulting the HF(B)
calculations, exact diagonalization (SM) and truncated SM calculations
(SM$^{mod}$) in selected $N=Z$ nuclei. See text for further details.\label{tab-occup}}
\begin{tabular}{ccccc}
\hline
Nucleus & orbital & HF(B) & SM & SM$^{mod}$ \\
\hline
$^{32}$S   & $0d_{5/2}$ & 6 & 5.48 & 6 \\   
           & $1s_{1/2}$ & 2 & 1.45 & 2 \\
           & $0d_{3/2}$ & 0 & 1.06 & 0 \\
\hline
$^{56}$Ni & $0f_{7/2}$ & 8 & 6.98 & 8 \\
          & $1p_{3/2}$ & 0 & 0.46 & 0 \\
          & $0f_{5/2}$ & 0 & 0.48 & 0 \\
          & $1p_{1/2}$ & 0 & 0.07 & 0 \\
\hline
$^{104}$Te & $0g_{7/2}$ & 0.45 & 0.53 & 0.41 \\
           & $1d_{5/2}$ & 0.95 & 0.90 & 1.02 \\
           & $2s_{1/2}$ & 0.36 & 0.29 & 0.36 \\
           & $1d_{3/2}$ & 0.24 & 0.16 & 0.21 \\
           & $0h_{11/2}$& 0.0  & 0.12 & 0.0 \\
\hline 
$^{108}$Xe &  $0g_{7/2}$ & 1.40 & 1.37 & 1.31 \\
	   &  $1d_{5/2}$ & 1.58 & 1.49 & 1.67 \\
	   &  $2s_{1/2}$ & 0.55 & 0.53 & 0.64 \\
	   &  $1d_{3/2}$ & 0.39 & 0.31 & 0.38 \\
	   &  $0h_{11/2}$& 0.08 & 0.30 & 0.0 \\
\hline
\end{tabular}
\end{table}

\begin{table}
\caption{Properties of the $M1$ strength distributions obtained 
in QRPA, SM and with modified SM wave-functions of the ground state, see text for further details. \label{tab-mod}}
\begin{tabular}{ccccc}
\hline
Nucleus &   & QRPA & SM & SM$^{mod}$ \\
\hline
$^{32}$S & $S_0$     &   8.21   & 5.68    & 10.55\\
         & $\bar S$  &   11.07  & 12.36   & 11.41\\
         & $\Delta S$&    0.27  & 3.20    & 0.30\\   
\hline
$^{56}$Ni& $S_0$     &  12.59 & 11.68 & 15.06 \\
         & $\bar S$  &  10.82 & 11.58 & 10.99 \\
         & $\Delta S$&  0.23  & 2.66  & 0.24   \\
\hline
$^{104}$Te & $S_0$     & 3.31 & 3.81 & 3.27 \\
           & $\bar S$  & 3.09 & 5.10 & 4.03 \\
           & $\Delta S$& 0.58 & 1.96 & 0.95 \\
\hline
$^{108}$Xe & $S_0$     & 3.64 & 6.29 & 4.91 \\
           & $\bar S$  & 3.15 & 5.65 & 4.12\\
           & $\Delta S$& 0.98 & 2.08 & 1.43\\
\hline
\end{tabular}
\end{table}

\subsection{Role of pairing correlations \label{sec-pair}}
\begin{figure}
\includegraphics[width=0.4\textwidth]{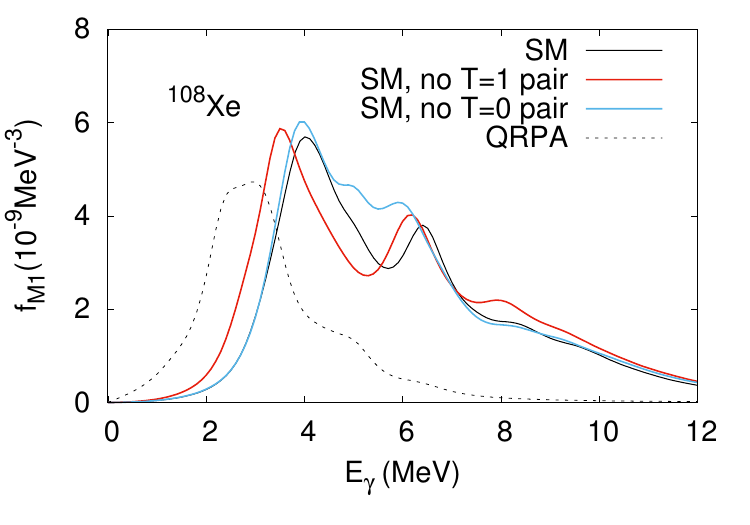}
\caption{$M1$ strength distributions of $^{108}$Xe obtained by exact diagonalization with the full SM Hamiltonian (SM)
and after removing schematic $T=1$ (no $T=1$ pair) and $T=0$ (no $T=0$ pair) interactions compared to the QRPA results. See text for details.\label{fig-xe108pn}}
\end{figure}

Since our selection contains many $N=Z$ nuclei, a remark about proton-neutron pairing correlations is in order. 
Those are not taken explicitly
into account in the mean-field calculations which constitutes a difference with respect to the SM diagonalization.
The role of $T=1$ and $T=0$ pairing interactions on rotational properties of lightest Xe nuclei
with GCN5082 interaction employed here was previously discussed within the shell model 
in Ref. \cite{xenon}. The deuteron-like $J=1$ isoscalar pairs were shown to
have a negligible presence in these nuclei and removing the $T=0$ pairing interaction did not affect the quadrupole properties.
In particular, the possibility of existence of the $T=0$ pair condensate in the ground state of $^{108}$Xe was refuted. 
The removal of isovector $T=1$ pairing was shown to impact mostly the moment of inertia without considerably alternating of the  
decay properties of the band. Here we repeat the calculations from Ref. \cite{xenon} to study the impact of p-n pairing interactions
on $M1$ distributions in $^{108}$Xe. To this end, a schematic pairing Hamiltonian was constructed with a strength adjusted to that
of the GCN5082 interaction on the two-body level. Further such a pairing Hamiltonian was substracted from the interaction and
the diagonalization of the $0^+$ state carried out, followed by a calculation of the strength function.
Figure \ref{fig-xe108pn} shows shell-model results with the full GCN5082 Hamiltonian
and after removal of the $T=0$ and $T=1$ schematic pairing interactions.  
The $T=0, J=1$ proton-neutron pairing interaction does not play major role:
the binding energy of the ground state is higher by 640keV and the sum rule is enlarged by $8\%$ without those correlations. 
The removal of the $T=1, J=0$ interactions has a bigger, though still limited impact, lowering the binding of the $0^+$ by 900 keV 
and increasing the total strength by $11\%$. As can be seen in the figure, once convoluted with Lorentzians, the 
distributions look fairly similar: the whole distribution is shifted down when the $T=1$ pairing is absent but the shape  
remains the same as in the full calculation. The absence of the $T=0$ pairing produces no effect at the lowest energies but  
more strength is accumulated around 5MeV. Overall, these effects are not significant enough to explain the difference with QRPA. The centroids of the 3 distributions agree within 100keV and the widths within 300keV. 
This little influence of pairing in the shell-model calculation of $^{108}$Xe
is not astonishing as its structure, similarly to the structure of many other
nuclei along the $N=Z$ line, is dominated by quadrupole correlations in the shell-model picture. 
The proton-neutron pairing correlations in the $N=Z$ nuclei studied here are thus of minor importance,
and one can suppose that taking them into account on the HFB level would not cure the 
rather important model differences. 

\begin{figure}
\includegraphics[width=0.4\textwidth]{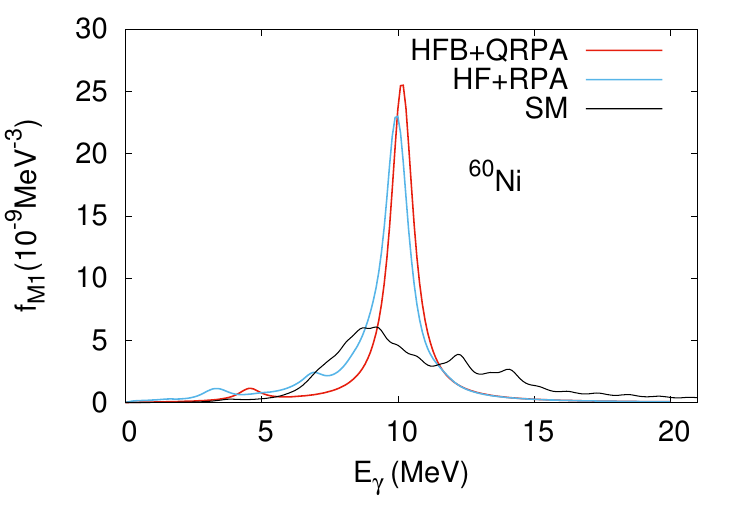}
\includegraphics[width=0.4\textwidth]{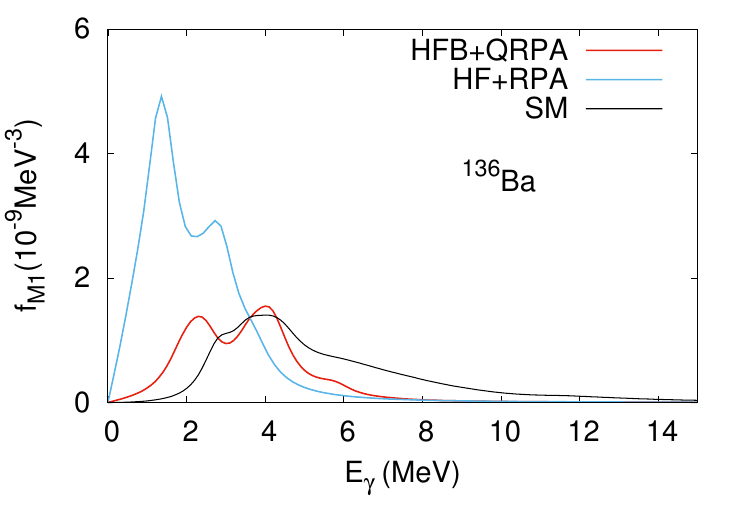}
\caption{Comparison of the HF+RPA, HFB+QRPA and SM strengths in $^{60}$Ni and $^{136}$Ba.\label{fig-pair}}
\end{figure}

Now let us turn back to the $T=1$ pairing correlations and their role in the QRPA calculations.  
As said before, the results of Ref. \cite{Stetcu2003} exhibited similar, systematic behaviors
of the centroids and widths of the computed strength functions as we observe here
for the magnetic dipole. This previous study was done within the RPA method only and thus pointed to the 
pairing correlations as possibly improving the results. 
To illustrate the effects of pairing in more detail,
we have computed ${}^{60}$Ni and $^{136}$Ba nuclei using HF+RPA approach and compared to HFB+QRPA results, 
as depicted in Fig. \ref{fig-pair}. 
Clearly, the presence of pairing correlations is responsible for a shift of the strength of around 2MeV in ${}^{136}$Ba that is
due to the lower energy of the HFB vacuum compared to the HF one. Pairing correlations also help with the spreading of the strength that turns out to be more fragmented. A shift of the centroid in the right direction is also observed in ${}^{60}$Ni although less pronounced than in ${}^{136}$Ba. This is probably explained by the fact that ${}^{60}$Ni is only singly open-shell and only the neutrons are paired in the HFB calculation. These results suggest that symmetry-restored QRPA calculations \cite{Federschmidt1985} (in which the mean-field is expected to be more paired) might give results closer to SM.

\subsection{Influence of triaxiality \label{sec-triax}}
\begin{figure}
\includegraphics[width=0.4\textwidth]{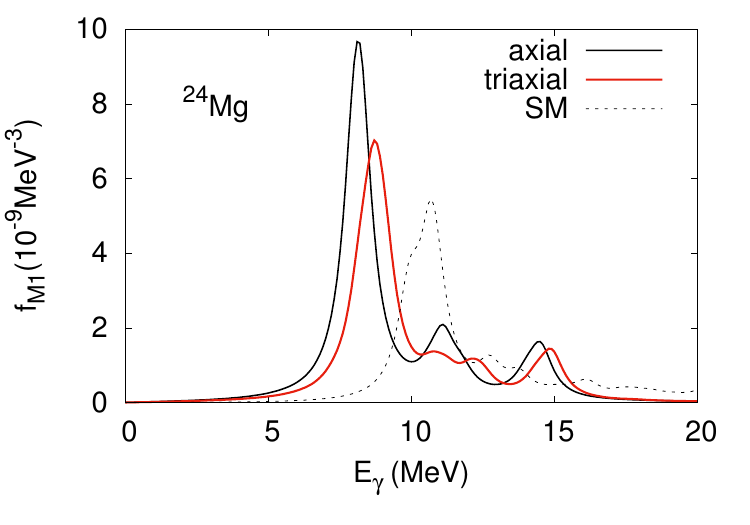}
\includegraphics[width=0.4\textwidth]{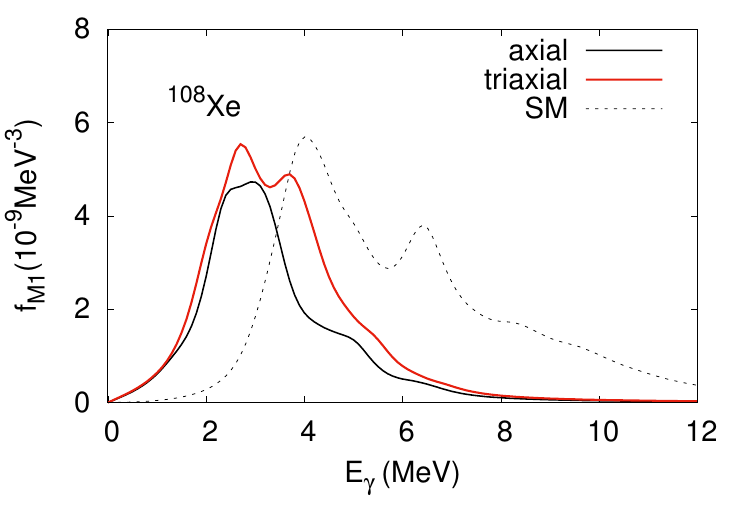}
\caption{Comparison of axially-deformed and triaxial QRPA calculations 
to exact diagonalization for $^{24}$Mg and $^{108}$Xe nuclei.\label{fig-triax}}
\end{figure}

In Fig. \ref{fig-triax} the QRPA results obtained starting from the axially-deformed
and triaxial mean-field solutions are shown in both triaxial
nuclei studied here, $^{24}$Mg and $^{108}$Xe. 
The changes due to triaxiality seem minor but go into the desired direction in both cases
(note that the behavior of axial and non-axial results is different in both nuclei):
In $^{24}$Mg the SM calculation gives $S_0$=4.24$\mu_N^2$,  $\bar S$=12.41MeV and $\Delta S$=3.75MeV.
The QRPA calculation based on the axially-deformed mean-field yields 
$S_0$=5.72$\mu_N^2$,  $\bar S$=9.39MeV and $\Delta S$=2.19MeV.
As can be noted from the figure, the inclusion of non-axiality in the ground state 
provides some reduction of the total strength ($S_0$=4.96$\mu_N^2$) and shifts the centroid 
to higher energies ($\bar S$=9.92MeV).
There is however no broadening of the distribution. Contrary to $^{24}$Mg, the total strength in $^{108}$Xe
is increased in the triaxial calculation from $S_0$=3.64$\mu_N^2$ to $S_0$=4.87$\mu_N^2$, bringing the solution 
to a slightly better agreement with the SM one:
$S_0$=6.29$\mu_N^2$. The centroid shifts by 200keV to the higher energy and is located at 3.34MeV, 
still being too low with respect to the SM value of 5.65MeV.

Similarily to what has been observed in Sec. \ref{sec-pair}, symmetry-breaking systematically
goes in the direction of SM, and further supports the idea that symmetry-restored QRPA might
potentially help by favoring large symmetry breaking in the reference state and therefore
improve prediction of $M1$ strength functions.

\section{Excited states: absorption and emission strength\label{sec-lee}}

Diverting our attention from the strength functions associated with the
nuclear ground state, we turn to the $M1$ photoabsorption and photoemission 
strength at finite excitation energy, focusing on the origin of the LEE and
whether or not it can be reproduced through QRPA calculations. We discuss 
first the photoabsorption and -emission strengths (and their difference) 
obtained from direct diagonalization, illustrating the presence of an LEE. 
We then extend our discussion to FT-QRPA: we compare exact and FT-QRPA results 
for both the absorption and emission strengths. To finish this section, we 
discuss future perspectives on the development of approaches that can account for
this physical effect and yet avoid the computational cost of exact diagonalization. 

\subsection{Exact diagonalization: absorption and emission}
\label{sec:SM}

\begin{figure}
\includegraphics[width=0.4\textwidth]{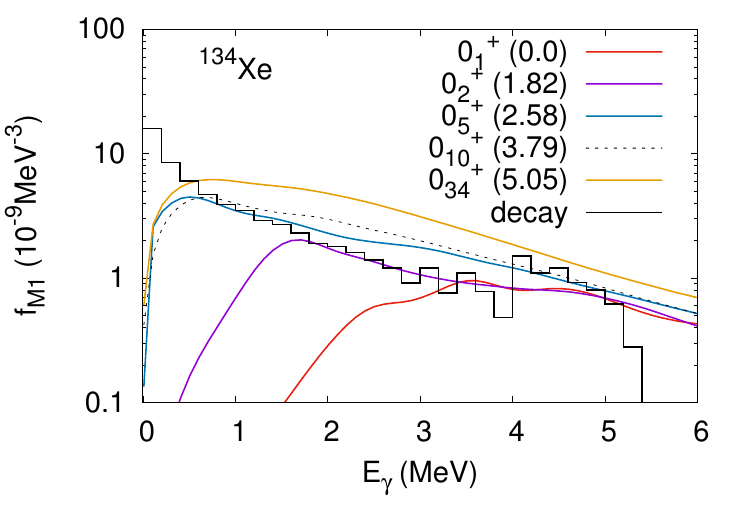}
\includegraphics[width=0.4\textwidth]{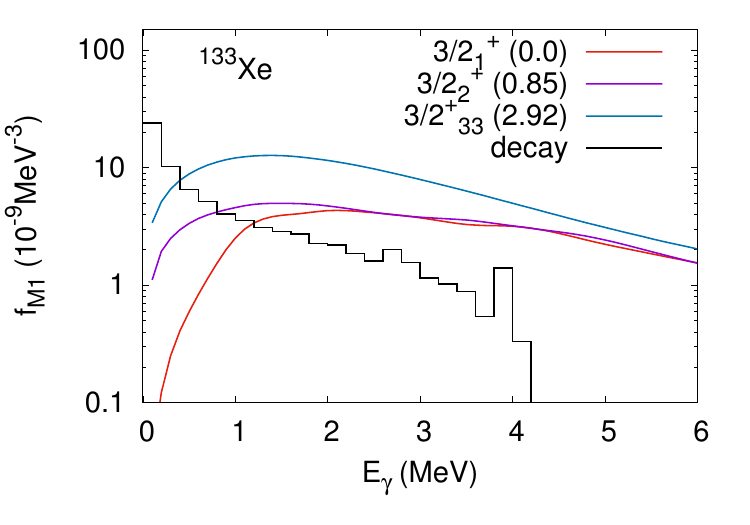}
\caption{Photoabsorption (colored, smooth lines) versus photoemission 
(black bins) strength functions obtained from exact diagonalization for
$^{134}$Xe (top panel) and $^{133}$Xe (bottom panel). We label the photoabsorption 
curves with the quantum number and excitation energy (in parentheses, expressed
in MeV) of the corresponding excited state.}
\label{fig-xe}
\end{figure}

In Fig.~\ref{fig-xe}, we compare the photoabsorption and emission strengths obtained
from exact diagonalization in two heavy nuclei, $^{134}$Xe (top panel) and 
$^{133}$Xe (bottom panel). The figure includes the absorption strength 
for the ground state and several excited states, indicated by their quantum
number and excitation energy. The decay strengths were computed by averaging 
transitions from many excited states using Eq. \ref{eq-fm1}, including all 
excited states up to 6.0MeV and $J=7$ for $^{134}$Xe and up to 4.0MeV and $J=15/2$. 
This selection included states of both parities for $^{133}$Xe, but we limited
ourselves to positive parity states for the even-even nucleus for the sake
of comparison to FT-QRPA results.

The LEE is clearly evident for $^{134}$Xe: the deexcitation strength in the bin
of lowest $E_{\gamma}$ (0-0.2 MeV) is the largest across the entire energy range. 
The absorption strength of the nucleus grows with increasing excitation energy
and approximates the decay strength across almost the entire range of the 
figure, except for the very lowest $\gamma$-ray energy bin. This figure 
illustrates the origin of the LEE as discussed in preceding shell model 
studies: the LEE consists of low-energy $\gamma$-transitions connecting the 
excited states in the quasi-continuum of nuclear levels. These conclusions 
are not significantly affected by our selection of states to compute
the decay strengths~\cite{Sieja2018}: (i) they are only weakly dependent on the 
considered spin and excitation energy range and (ii) negative parity states
in those nuclei contribute even more to the decay strength at low energy, leading 
to an even more pronounced LEE for $^{134}$Xe, had we considered them. In 
fact, even restricting the calculation to $0^+$ and $1^+$ excited states 
still leads to similar shape and magnitude of the decay strength. 

The deexcitation strength of the odd-even nucleus $^{133}$Xe is qualitatively
similar to that of $^{134}$Xe, taking into account the 4 MeV cutoff in excitation
energy in our calculation. There is a qualitative
difference in the photoabsorption strengths however: in contrast to the even-even 
nucleus  the odd-mass nucleus has significant strength for $E_{\gamma}$ below 1 MeV even for the ground state.
The origin of this difference is pairing: the low-energy spectrum of the even-even nucleus
is much more sparse than that of its odd-mass neighbour, with the first $1^+$ 
in $^{134}$Xe at 2MeV and the first excited state $1/2^+$ in $^{133}$Xe at 0.25MeV.

\subsection{QRPA at finite temperature: absorption}

\begin{figure}
\includegraphics[width=0.45\textwidth]{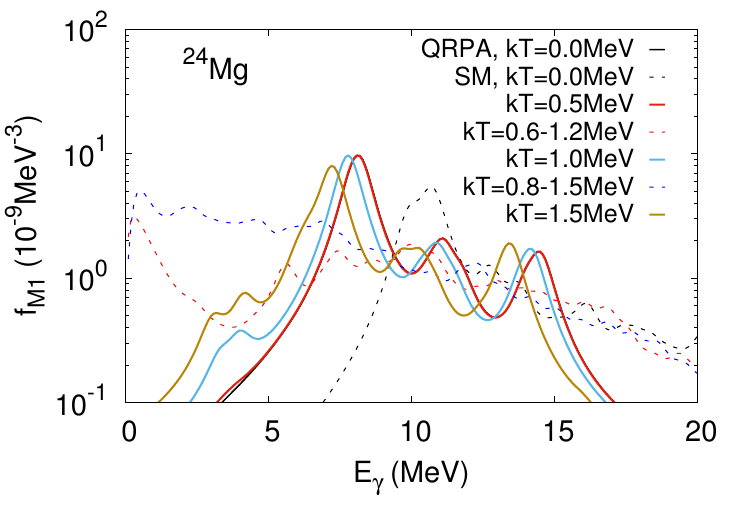}
\includegraphics[width=0.45\textwidth]{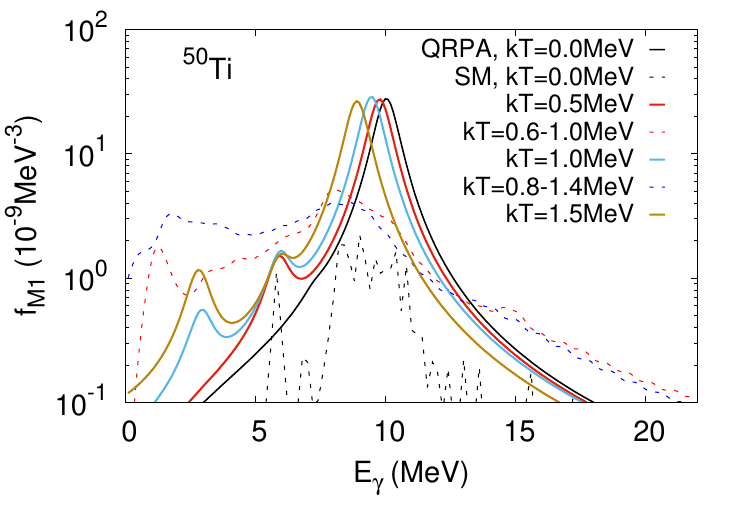}
\includegraphics[width=0.45\textwidth]{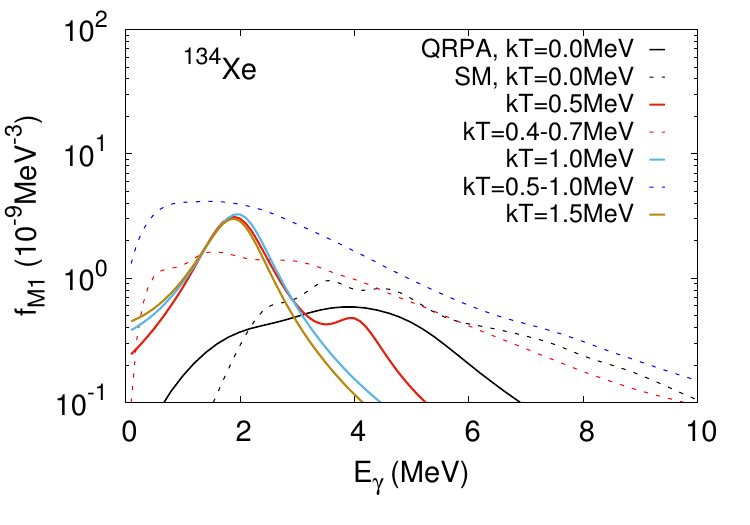}
\caption{
The $M1$ photoresponse strength functions for $^{24}$Mg (top), $^{50}$Ti (middle)
and $^{134}$Xe (bottom) as obtained through FT-QRPA calculations for different values of the temperature (solid lines) and through
exact diagonalization from initial states of different excitation energy 
(dashed lines).}
 \label{fig-qrpa-abs}
\end{figure}

We first discuss the evolution of the $M1$ photoabsorption strength function 
with increasing excitation energy: Fig.~\ref{fig-qrpa-abs} shows the (FT-)~QRPA
strength functions for different values of the temperature in solid lines for one 
nucleus in each of the model spaces. We remind the reader that, as discussed
in Sec.~\ref{sec:qrpa}, the photoabsorption $f_{\rm M1}$ strengths are obtained 
from the positive-energy part of the FT-QRPA microscopic strength function $S_{\rm M1}$.
For each nucleus, an increase in temperature shifts the centroid to slightly 
lower $\gamma$-ray energies. The total strength obtained rises initially when 
increasing the temperature, but this trend reverses at the highest temperatures 
due to the limitations of the model space. Approaches based on energy density 
functionals are typically not limited to valence spaces: we anticipate that in 
such approaches the total strength will monotonously rise with 
increasing temperature.

\begin{figure}
\centering
\includegraphics[width=.45\textwidth]{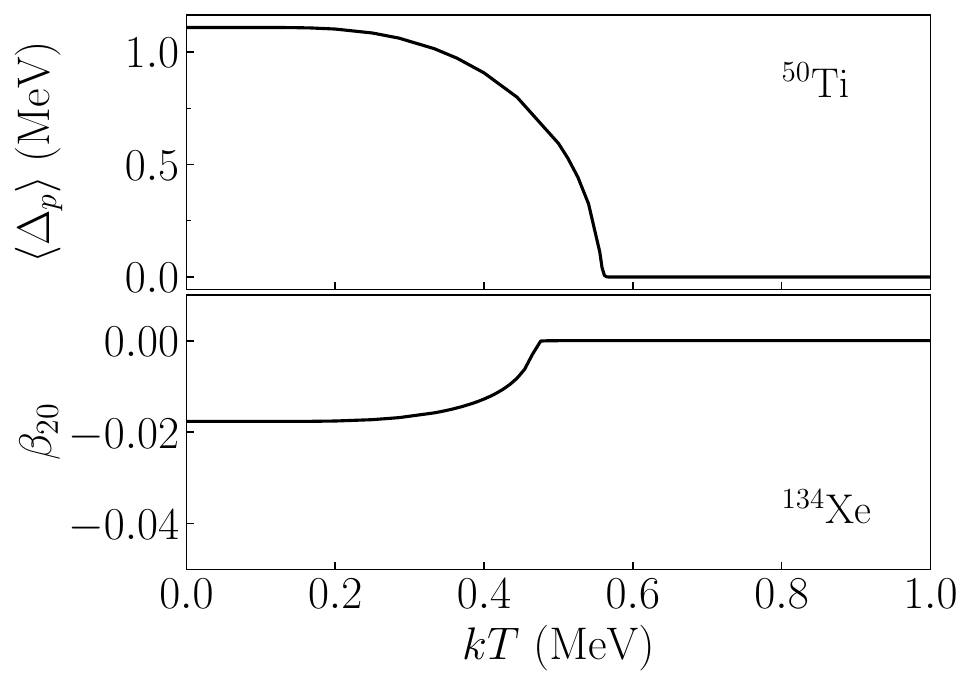}
\caption{ Illustration of the temperature phase transitions for the FT-QRPA. 
          Top panel: average proton pairing gap in $^{50}$Ti as a function of 
          $kT$. Bottom panel: quadrupole deformation $\beta_{20}$ for $^{134}$Xe.}
\label{fig:transition}
\end{figure}

Aside from these effects that affect the strength function as a function of
temperature in a smooth way, there are discontinuous changes to be 
seen in the middle and bottom panels of Fig.~\ref{fig-qrpa-abs}. For $^{50}$Ti, 
this is the development of two additional peaks, first near 5 MeV and at 
higher temperatures also near 2.5 MeV. For $^{134}$Xe, the change in shape of
the absorption strength is even more dramatic. In both cases, these changes
reflect the discontinuous structural changes in the underlying mean-field
solution that mark a temperature phase transition. These are illustrated in 
Fig.~\ref{fig:transition}: the top panel shows the pairing phase transition
in $^{50}$Ti by means of the average proton pairing gap while the bottom panel 
shows the shape transition in $^{134}$Xe by means of its quadrupole deformation
$\beta_{20}$. $^{24}$Mg also undergoes a phase transition from a prolate
to a spherical shape, but for our model space and Hamiltonian this occurs for 
temperatures above those we consider here~\cite{HF-SHELL}.

It is not trivial to compare FT-QRPA and SM results: the 
former depend on temperature and the latter on excitation energy. We relate the 
temperature of an excited 
$0^{+}$ state to its excitation energy through a (phenomenological) model of 
the level density of the corresponding nucleus: $T=\sqrt{(E^*-\delta)/a}$ with 
$E^*$ the calculated
excitation energy, $\delta$ a pairing energy shift and $a$ the level density parameter.
We use values of the latter two parameters from both the back-shifted Fermi gas 
and Gilbert-Cameron model as tabulated in Ref.~\cite{RIPL-3}, resulting in two
temperatures for each excited SM state that we take as an indicative range.
The resulting $M1$ strength functions obtained through exact diagonalization 
are drawn in Fig.~\ref{fig-qrpa-abs} as dashed lines. The centroids of the SM 
$M1$ absorption strength for excited $0^+$ states 
shift to energies that are several MeV lower than that of the ground state: 
many more $1^+$ states find themselves in the direct vicinity of excited states.
The total strength rises monotonously with temperature in this range of 
$E_{\gamma}$, although also the diagonalization approach will eventually face the limitations of the valence space at even higher excitation energies. 

Comparing FT-QRPA and SM, we see a qualitative
similarity in that the centroids shift to lower $E_{\gamma}$ and that the 
total strengths increase with increasing excitation energy in both approaches.
It is however immediately clear from all panels in Fig.~\ref{fig-qrpa-abs}
that these effects are too small in FT-QRPA: the 
FT-QRPA absorption strength at high excitation energy differs dramatically from
the exact result in all cases. This is in spite of the rather fair reproduction of the 
ground state absorption strength for all three nuclei in Fig.~\ref{fig-qrpa-abs}, 
although the overall performance of FT-QRPA improves somewhat
for heavier nuclei.

The reason for the failure of FT-QRPA is its level density: because it is 
limited to two-quasiparticle excitations, the total number of many-body states 
that can be constructed is much smaller than those in an exact diagonalization. 
This is not so problematic when studying photoabsorption of the nuclear ground 
state, as many of the missing states are located at (comparatively) high 
excitation energy. Although the introduction of finite temperature allows for 
the construction of additional many-body states compared to a ground state 
calculation (the thermal unblocking effect referred to in Sec.~\ref{sec:qrpa}), 
this does not suffice to capture the complexity of the entire many-body space.
As an illustration: only 48 1p-1h excitations with $J^\pi=1^+$ can be constructed for $^{24}$Mg in the $sd$-shell while there are in total 3096 $1^+$ states that figure in an exact diagonalization. 

Although our comparison is limited to even-even nuclei, we note that it is likely
that FT-QRPA would compare somewhat better to the SM result for odd-mass 
and odd-odd nuclei. In those, the level density at low excitation energy is
much higher as discussed in Sec.~\ref{sec:SM} such that FT-QRPA could possibly
be able to capture a part of the absorption strength at low $E_{\gamma}$ of the
exact results.

\subsection{QRPA at finite temperature: deexcitation}

\begin{figure}
\includegraphics[width=0.45\textwidth]{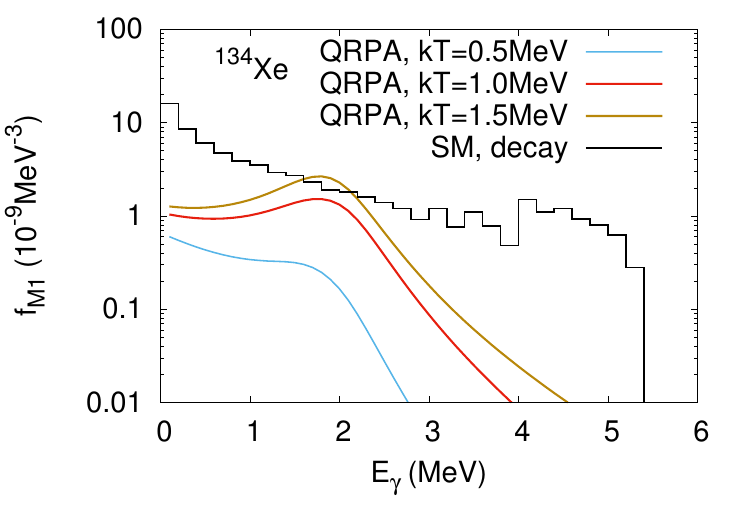}
\caption{Photoemission strength functions in $^{134}$Xe obtained in the 
FT-QRPA approach versus the deexcitation strength function obtained from the 
shell-model diagonalization (see text for details). \label{fig-decay}}
\end{figure}

In Fig.~\ref{fig-decay} we show the deexcitation strength function of $^{134}$Xe
obtained from SM and FT-QRPA calculations, the latter of which is derived from 
the microscopic strength function $S_{\rm M1}(E)$ at negative energy as discussed
in Sec.~\ref{sec:qrpa}\footnote{We remind the reader that $S_{\rm M1}(E)$ vanishes
at negative energy at $kT = 0$: the formalism reduces to QRPA and 
the nuclear ground state cannot decay by emission of a photon.}. 
The evaluation of the SM strength included excited states 
up to $\sim$6~MeV, which should correspond to a maximal temperature of $kT=0.6-1.1$MeV: 
we report FT-QRPA strength for a corresponding range of $kT=$0.5-1.5MeV. 

Looking first at the FT-QRPA results by themselves, we see a significant evolution 
in the overall decay strength: as the temperature increases from $kT = 0.5$ to $1.5$ MeV
the total strength increases significantly across the range of $E_{\gamma}$
\footnote{There is no discontinuous change due a phase transition visible on 
Fig.~\ref{fig-decay}; the shape phase transition for $^{134}$Xe is slightly
below $kT=0.5$ MeV.}. This increase is more rapid at higher $E_{\gamma}$, such that 
the centroid of the decay strength increases with increasing temperature. The 
nuclear susceptibility $\chi(\omega)$ depends on $kT$ implicitly and carries 
most of the structural information of the mean-field configuration, but it varies very slowly with 
increasing temperature in the absence of phase transitions~\cite{lipparini2003}. 
Most of the thermal evolution visible in Fig.~\ref{fig-decay} is due to the thermal 
prefactor in Eq.~\eqref{eq:prefact} which varies quickly with $kT$\footnote{This thermal prefactor was not included in the study of dipole
          response based on ab initio Hamiltonians of Ref.\cite{Beaujeault2022}. 
          The authors compared the susceptibility $\chi(\omega)$ with experimental data 
          on the photo-decay of $^{56}$Fe.}.

Although some temperature enhancement is visible, the FT-QRPA strength at the 
lowest $\gamma$-ray energies differs from the SM result by roughly an order of 
magnitude even at high temperature. As the temperature increases, the 
level density accessible to FT-QRPA enlarges but this effect is not sufficient 
to produce an LEE that is comparable to the one obtained from exact diagonalization.
Although some decay strength is produced at low energy, such strength
is the tail of a peak located at roughly 2 MeV; this is a generic feature of
even-even nuclei and it thus seems unlikely that FT-QRPA or other extensions 
of QRPA techniques that do not explicitly consider excited states would ever be
able to produce a sufficiently large LEE. Similarly, approaches that obtain 
decay strength functions through photoabsorption strength functions of the 
ground states in even-even nuclei will likely fail to produce the LEE, barring 
explicit inclusion through phenomenology. As discussed above, it is possible
the situation is less dire for odd-mass and odd-odd nuclei where the level
density at low energy is much higher.

As we mentioned above, the FT-QRPA is arguably the simplest extension of QRPA 
that directly results in decay strength functions. By virtue of the comparative 
simplicity of QRPA, FT-QRPA is likely the least numerically intensive many-body 
technique that provides such access. It is clear that this approach is however 
too simple to produce an LEE comparable to the SM result; one should look elsewhere
for an approach that can cover the nuclear chart but still provide accurate dipole
strength functions. There are multiple candidates for such an approach: a first 
is extending the QRPA framework with the calculation of all transitions between
pairs of constructed states as opposed to just those rates involving the nuclear
ground state. Using the excitation energies and the reduced matrix elements 
obtained in this way, it should be possible to construct a decay strength function
that probes a far larger level density although this has not yet been demonstrated at scale~\cite{Gaudefroy2018}.
Multi-phonon approaches constitute another path: these couple together multiple
QRPA phonons and overcome some of the downsides of QRPA such as the fragmentation
of the $M1$ strength.~\cite{tsoneva2008,Rusev2013}, but have so far not been
used to investigate the LEE due to $M1$ radiation. Although it has not been 
applied so far to describe dipole strength functions, the Projected Generator 
Coordinate Method (PGCM) is perhaps the most promising future avenue. PGCM has the potential 
to improve on (FT-)~QRPA by (i) restoring quantum numbers lost by spontaneous
symmetry breaking and hence enriching the nuclear ground state with additional
correlations and (ii) capturing vibrational and rotational degrees of freedom
within one single framework. Although this method scales modestly compared to exact 
diagonalization, polynomial as opposed to combinatorial, it remains demanding 
and has so far not been deployed at scale~\cite{duguet2022}. Finally, one 
could conceivably look for different computational techniques while keeping the 
power of SM results: Shell Model Monte Carlo (SMMC) techniques can provide exact results, 
up to statistical errors, in larger valence spaces than traditional diagonalization 
methods~\cite{Koonin1997}. Ref.~\cite{fanto2021} recently proved that one can 
use such techniques to study the LEE  in rare earth nuclei and it is plausible
that even heavier nuclei could be addressed, though it is unlikely that systematic
SMMC calculations will ever become feasible since the method remains bound to 
the construction of a valence space and effective interaction.

Although we conclude that standard (FT-)~QRPA is not able to describe the LEE 
of magnetic dipole strength functions, we mention here that the method will likely
fare better for the \emph{electric} dipole strength function. SM calculations
have shown that 1p-1h excitations typically suffice to describe low-energy
$E1$ strength~\cite{Sieja-PRL,Shimizu-Ca48} while multi-phonon approaches 
indicate that this strength can be captured by one-phonon calculations~\cite{Tsoneva2004}.
Because of this, we expect that most of the physically relevant part of the 
many-body space is accessible to (FT-)~QRPA and the corresponding description 
of $E1$ transitions to be more successful. We will test this expectation in a
forthcoming study along the same lines as this one. 

\section{Conclusions\label{sec-pro}}

We have compared the absorption and decay magnetic dipole strength functions obtained from
(FT-)~QRPA and exact diagonalization in identical shell-model valence spaces and
employing the same Hamiltonians. Our study spanned 25 even-even nuclei, from 
nuclei with $A \sim 28$ in the $sd$-shell up to medium-heavy nuclei with $A\sim 130$. 
Future work will be devoted to the study of electric dipole transitions along the same lines.
The ground state photo-absorption strength obtained from QRPA calculations is 
rather satisfactory: it typically agrees with the exact result within about 
20\%, although larger deviations occurred in our calculations. Other aspects
of the (FT-)~QRPA predictions are less appealing, and our results highlight 
two issues of this approach to obtain magnetic dipole strength functions. 

The first issue concerns the lack of correlations in the nuclear 
ground state, which causes a systematic shift of the centroids of the $M1$ strengths 
towards lower $\gamma$-ray energies. We established that the size of this
effect is somewhat lessened when the mean-field reference state incorporates
more correlations through spontaneous symmetry breaking: both the appearance
of triaxial deformation and pairing condensate tend to improve the agreement
with exact results, although we found proton-neutron pairing to be of very
limited relevance. Nevertheless, it is unlikely that even the most 
general symmetry-broken configurations will be able to completely offset this 
effect, but our observation indicates that symmetry-restoration techniques 
are a promising route since they tend to drive the reference state towards
less symmetrical configurations.

A second problem of the (FT-)~QRPA is the truncation of many-body space to 
two-quasiparticle excitations on top of a mean-field reference state. This 
approximation leads to a level density that is too low, even at high excitation 
energies, leading to (i) a lack of fragmentation of the strength and (ii) a lack of 
strength at low $E_{\gamma}$ for both photo-absorption and -decay strength 
functions of excited states. Most studies in the literature deal with the 
ground-state photo-absorption strength function of even-even nuclei; in this 
regime these deficiencies of QRPA are not immediately apparent since (i) a 
phenomenological smearing factor is incorporated to provide fragmentation and 
(ii) the sparsity of the low-energy spectrum of even-even nuclei forbids 
finite strength at low $E_{\gamma}$. The generalization of the formalism to
finite temperature slightly enlarges the model space and allows for the appearance
of some low-lying $M1$ strength, the effect is typically far too small compared to 
the exact result. In particular, this indicates that traditional (FT-)~QRPA approaches
may not be reliable to predict the presence or absence of a low-energy enhancement 
of the $M1$ strength below the neutron emission threshold. We discussed briefly
alternative approaches to tackle this issue in a more global way; development 
of a framework to extract $M1$ strength functions from PGCM calculations 
based on shell-model Hamiltonians is underway \cite{Bofos}.

\section*{Acknowledgments}
We gratefully acknowledge useful discussions with Sophie Péru.
W.R. is a Research Associate of the F.R.S.-FNRS.



\begin{thebibliography}{76}
\expandafter\ifx\csname natexlab\endcsname\relax\def\natexlab#1{#1}\fi
\expandafter\ifx\csname bibnamefont\endcsname\relax
  \def\bibnamefont#1{#1}\fi
\expandafter\ifx\csname bibfnamefont\endcsname\relax
  \def\bibfnamefont#1{#1}\fi
\expandafter\ifx\csname citenamefont\endcsname\relax
  \def\citenamefont#1{#1}\fi
\expandafter\ifx\csname url\endcsname\relax
  \def\url#1{\texttt{#1}}\fi
\expandafter\ifx\csname urlprefix\endcsname\relax\def\urlprefix{URL }\fi
\providecommand{\bibinfo}[2]{#2}
\providecommand{\eprint}[2][]{\url{#2}}

\bibitem[{\citenamefont{Arnould and Goriely}(2020)}]{Arnould2020}
\bibinfo{author}{\bibfnamefont{M.}~\bibnamefont{Arnould}} \bibnamefont{and}
  \bibinfo{author}{\bibfnamefont{S.}~\bibnamefont{Goriely}},
  \bibinfo{journal}{Progress in Particle and Nuclear Physics}
  \textbf{\bibinfo{volume}{112}}, \bibinfo{pages}{103766}
  (\bibinfo{year}{2020}), ISSN \bibinfo{issn}{01466410},
  \bibinfo{note}{arXiv:2001.11228 [astro-ph, physics:nucl-th]},
  \urlprefix\url{http://arxiv.org/abs/2001.11228}.

\bibitem[{\citenamefont{Feshbach et~al.}(1954)\citenamefont{Feshbach, Porter,
  and Weiskopf}}]{Hauser-Feshbach}
\bibinfo{author}{\bibfnamefont{H.}~\bibnamefont{Feshbach}},
  \bibinfo{author}{\bibfnamefont{C.}~\bibnamefont{Porter}}, \bibnamefont{and}
  \bibinfo{author}{\bibfnamefont{V.}~\bibnamefont{Weiskopf}},
  \bibinfo{journal}{Phys. Rev. C} \textbf{\bibinfo{volume}{96}},
  \bibinfo{pages}{448} (\bibinfo{year}{1954}).

\bibitem[{SLO()}]{SLO}
\bibinfo{howpublished}{Handbook for calculations of nuclear reaction
  data.RIPL.IAEA- TECDOC-1034, August 1998; http: //www-nds.iaea.or.at /ripl/}.

\bibitem[{RIP()}]{RIPL-3}
\bibinfo{howpublished}{https://www-nds.iaea.org/RIPL-3/}.

\bibitem[{osl()}]{oslo-website}
\bibinfo{howpublished}{http://www.mn.uio.no/fysikk/}.

\bibitem[{\citenamefont{Larsen et~al.}(2013)}]{Larsen-Fe56}
\bibinfo{author}{\bibfnamefont{A.~C.} \bibnamefont{Larsen}}
  \bibnamefont{et~al.}, \bibinfo{journal}{Phys. Rev. Lett.}
  \textbf{\bibinfo{volume}{111}}, \bibinfo{pages}{242504}
  (\bibinfo{year}{2013}).

\bibitem[{\citenamefont{Utsunomiya et~al.}(2013)\citenamefont{Utsunomiya,
  Goriely, Kondo, Iwamoto, Akimune, Yamagata, Toyokawa, Harada, Kitatani, Lui
  et~al.}}]{enh-Mo}
\bibinfo{author}{\bibfnamefont{H.}~\bibnamefont{Utsunomiya}},
  \bibinfo{author}{\bibfnamefont{S.}~\bibnamefont{Goriely}},
  \bibinfo{author}{\bibfnamefont{T.}~\bibnamefont{Kondo}},
  \bibinfo{author}{\bibfnamefont{C.}~\bibnamefont{Iwamoto}},
  \bibinfo{author}{\bibfnamefont{H.}~\bibnamefont{Akimune}},
  \bibinfo{author}{\bibfnamefont{T.}~\bibnamefont{Yamagata}},
  \bibinfo{author}{\bibfnamefont{H.}~\bibnamefont{Toyokawa}},
  \bibinfo{author}{\bibfnamefont{H.}~\bibnamefont{Harada}},
  \bibinfo{author}{\bibfnamefont{F.}~\bibnamefont{Kitatani}},
  \bibinfo{author}{\bibfnamefont{Y.-W.} \bibnamefont{Lui}},
  \bibnamefont{et~al.}, \bibinfo{journal}{Phys. Rev. C}
  \textbf{\bibinfo{volume}{88}}, \bibinfo{pages}{015805}
  (\bibinfo{year}{2013}).

\bibitem[{\citenamefont{B\"urger et~al.}(2012)}]{end-Sc43}
\bibinfo{author}{\bibfnamefont{A.}~\bibnamefont{B\"urger}}
  \bibnamefont{et~al.}, \bibinfo{journal}{Phys. Rev. C}
  \textbf{\bibinfo{volume}{85}}, \bibinfo{pages}{064328}
  (\bibinfo{year}{2012}).

\bibitem[{\citenamefont{Larsen et~al.}(2012)}]{Larsen-Ti44}
\bibinfo{author}{\bibfnamefont{A.~C.} \bibnamefont{Larsen}}
  \bibnamefont{et~al.}, \bibinfo{journal}{Phys. Rev. C}
  \textbf{\bibinfo{volume}{85}}, \bibinfo{pages}{014320}
  (\bibinfo{year}{2012}).

\bibitem[{\citenamefont{Larsen and Goriely}(2010)}]{Larsen2010}
\bibinfo{author}{\bibfnamefont{A.~C.} \bibnamefont{Larsen}} \bibnamefont{and}
  \bibinfo{author}{\bibfnamefont{S.}~\bibnamefont{Goriely}},
  \bibinfo{journal}{Phys. Rev. C} \textbf{\bibinfo{volume}{82}},
  \bibinfo{pages}{014318} (\bibinfo{year}{2010}).

\bibitem[{\citenamefont{Schwengner et~al.}(2013)\citenamefont{Schwengner,
  Frauendorf, and Larsen}}]{Schwengner-Mo}
\bibinfo{author}{\bibfnamefont{R.}~\bibnamefont{Schwengner}},
  \bibinfo{author}{\bibfnamefont{S.}~\bibnamefont{Frauendorf}},
  \bibnamefont{and} \bibinfo{author}{\bibfnamefont{A.~C.}
  \bibnamefont{Larsen}}, \bibinfo{journal}{Phys. Rev. Lett.}
  \textbf{\bibinfo{volume}{111}}, \bibinfo{pages}{232504}
  (\bibinfo{year}{2013}).

\bibitem[{\citenamefont{Sieja}(2017{\natexlab{a}})}]{Sieja-PRL}
\bibinfo{author}{\bibfnamefont{K.}~\bibnamefont{Sieja}},
  \bibinfo{journal}{Phys. Rev. Lett.} \textbf{\bibinfo{volume}{119}},
  \bibinfo{pages}{052502} (\bibinfo{year}{2017}{\natexlab{a}}).

\bibitem[{\citenamefont{Schwengner et~al.}(2017)\citenamefont{Schwengner,
  Frauendorf, and Brown}}]{Schwengner-Fe}
\bibinfo{author}{\bibfnamefont{R.}~\bibnamefont{Schwengner}},
  \bibinfo{author}{\bibfnamefont{S.}~\bibnamefont{Frauendorf}},
  \bibnamefont{and} \bibinfo{author}{\bibfnamefont{B.~A.} \bibnamefont{Brown}},
  \bibinfo{journal}{Phys. Rev. Lett.} \textbf{\bibinfo{volume}{118}},
  \bibinfo{pages}{092502} (\bibinfo{year}{2017}).

\bibitem[{\citenamefont{Sieja}(2017{\natexlab{b}})}]{Sieja-EPJA}
\bibinfo{author}{\bibfnamefont{K.}~\bibnamefont{Sieja}}, \bibinfo{journal}{EPJ
  Web of Conferences} \textbf{\bibinfo{volume}{146}}, \bibinfo{pages}{05004}
  (\bibinfo{year}{2017}{\natexlab{b}}).

\bibitem[{\citenamefont{Karampagia et~al.}(2017)\citenamefont{Karampagia,
  Brown, and Zelevinsky}}]{Brown-V}
\bibinfo{author}{\bibfnamefont{S.}~\bibnamefont{Karampagia}},
  \bibinfo{author}{\bibfnamefont{B.~A.} \bibnamefont{Brown}}, \bibnamefont{and}
  \bibinfo{author}{\bibfnamefont{V.}~\bibnamefont{Zelevinsky}},
  \bibinfo{journal}{Phys. Rev. C} \textbf{\bibinfo{volume}{95}},
  \bibinfo{pages}{024322} (\bibinfo{year}{2017}).

\bibitem[{\citenamefont{Frauendorf et~al.}(2015)}]{LEMAR}
\bibinfo{author}{\bibfnamefont{S.}~\bibnamefont{Frauendorf}}
  \bibnamefont{et~al.}, \bibinfo{journal}{Eur. Phys. J.}
  \textbf{\bibinfo{volume}{93}}, \bibinfo{pages}{04002} (\bibinfo{year}{2015}).

\bibitem[{\citenamefont{Karampagia et~al.}(2018)\citenamefont{Karampagia,
  Brown, and Zelevinsky}}]{Karampagia2018}
\bibinfo{author}{\bibfnamefont{S.}~\bibnamefont{Karampagia}},
  \bibinfo{author}{\bibfnamefont{B.~A.} \bibnamefont{Brown}}, \bibnamefont{and}
  \bibinfo{author}{\bibfnamefont{V.}~\bibnamefont{Zelevinsky}},
  \bibinfo{journal}{Journal of Physics: Conf. Series}
  \textbf{\bibinfo{volume}{966}}, \bibinfo{pages}{012Â031}
  (\bibinfo{year}{2018}).

\bibitem[{\citenamefont{Midtb\o{} et~al.}(2018)\citenamefont{Midtb\o{}, Larsen,
  Renstr\o{}m, Bello~Garrote, and Lima}}]{Mitbo-LEE}
\bibinfo{author}{\bibfnamefont{J.~E.} \bibnamefont{Midtb\o{}}},
  \bibinfo{author}{\bibfnamefont{A.~C.} \bibnamefont{Larsen}},
  \bibinfo{author}{\bibfnamefont{T.}~\bibnamefont{Renstr\o{}m}},
  \bibinfo{author}{\bibfnamefont{F.~L.} \bibnamefont{Bello~Garrote}},
  \bibnamefont{and} \bibinfo{author}{\bibfnamefont{E.}~\bibnamefont{Lima}},
  \bibinfo{journal}{Phys. Rev. C} \textbf{\bibinfo{volume}{98}},
  \bibinfo{pages}{064321} (\bibinfo{year}{2018}).

\bibitem[{\citenamefont{Litvinova and Belov}(2013)}]{elena-Mo}
\bibinfo{author}{\bibfnamefont{E.}~\bibnamefont{Litvinova}} \bibnamefont{and}
  \bibinfo{author}{\bibfnamefont{N.}~\bibnamefont{Belov}},
  \bibinfo{journal}{Phys. Rev. C} \textbf{\bibinfo{volume}{88}},
  \bibinfo{pages}{031302} (\bibinfo{year}{2013}).

\bibitem[{\citenamefont{Wibowo and Litvinova}(2019)}]{Wibowo2019}
\bibinfo{author}{\bibfnamefont{H.}~\bibnamefont{Wibowo}} \bibnamefont{and}
  \bibinfo{author}{\bibfnamefont{E.}~\bibnamefont{Litvinova}},
  \bibinfo{journal}{Phys. Rev. C} \textbf{\bibinfo{volume}{100}},
  \bibinfo{pages}{024307} (\bibinfo{year}{2019}),
  \urlprefix\url{https://link.aps.org/doi/10.1103/PhysRevC.100.024307}.

\bibitem[{\citenamefont{Beaujeault-Taudi\`ere
  et~al.}(2023)\citenamefont{Beaujeault-Taudi\`ere, Frosini, Ebran, Duguet,
  Roth, and Som\`a}}]{Beaujeault2022}
\bibinfo{author}{\bibfnamefont{Y.}~\bibnamefont{Beaujeault-Taudi\`ere}},
  \bibinfo{author}{\bibfnamefont{M.}~\bibnamefont{Frosini}},
  \bibinfo{author}{\bibfnamefont{J.-P.} \bibnamefont{Ebran}},
  \bibinfo{author}{\bibfnamefont{T.}~\bibnamefont{Duguet}},
  \bibinfo{author}{\bibfnamefont{R.}~\bibnamefont{Roth}}, \bibnamefont{and}
  \bibinfo{author}{\bibfnamefont{V.}~\bibnamefont{Som\`a}},
  \bibinfo{journal}{Phys. Rev. C} \textbf{\bibinfo{volume}{107}},
  \bibinfo{pages}{L021302} (\bibinfo{year}{2023}),
  \urlprefix\url{https://link.aps.org/doi/10.1103/PhysRevC.107.L021302}.

\bibitem[{\citenamefont{Brown and Larsen}(2014)}]{brown-Fe56}
\bibinfo{author}{\bibfnamefont{B.~A.} \bibnamefont{Brown}} \bibnamefont{and}
  \bibinfo{author}{\bibfnamefont{A.~C.} \bibnamefont{Larsen}},
  \bibinfo{journal}{Phys. Rev. Lett.} \textbf{\bibinfo{volume}{113}},
  \bibinfo{pages}{252502} (\bibinfo{year}{2014}).

\bibitem[{\citenamefont{Ring and Schuck}(1980{\natexlab{a}})}]{Ring80}
\bibinfo{author}{\bibfnamefont{P.}~\bibnamefont{Ring}} \bibnamefont{and}
  \bibinfo{author}{\bibfnamefont{P.}~\bibnamefont{Schuck}},
  \emph{\bibinfo{title}{The nuclear many-body problem}}
  (\bibinfo{publisher}{Springer-Verlag, Berlin},
  \bibinfo{year}{1980}{\natexlab{a}}).

\bibitem[{\citenamefont{Paar et~al.}(2007)\citenamefont{Paar, Vretenar, Khan,
  and Colo}}]{Paar2007}
\bibinfo{author}{\bibfnamefont{N.}~\bibnamefont{Paar}},
  \bibinfo{author}{\bibfnamefont{D.}~\bibnamefont{Vretenar}},
  \bibinfo{author}{\bibfnamefont{E.}~\bibnamefont{Khan}}, \bibnamefont{and}
  \bibinfo{author}{\bibfnamefont{G.}~\bibnamefont{Colo}},
  \bibinfo{journal}{Rept. Prog. Phys.} \textbf{\bibinfo{volume}{70}},
  \bibinfo{pages}{691} (\bibinfo{year}{2007}).

\bibitem[{\citenamefont{Terasaki and Engel}(2010)}]{Terasaki2010}
\bibinfo{author}{\bibfnamefont{J.}~\bibnamefont{Terasaki}} \bibnamefont{and}
  \bibinfo{author}{\bibfnamefont{J.}~\bibnamefont{Engel}},
  \bibinfo{journal}{Phys. Rev. C} \textbf{\bibinfo{volume}{82}},
  \bibinfo{pages}{034326} (\bibinfo{year}{2010}),
  \urlprefix\url{https://link.aps.org/doi/10.1103/PhysRevC.82.034326}.

\bibitem[{\citenamefont{Martini
  et~al.}(2016{\natexlab{a}})\citenamefont{Martini, P{\'{e} }ru, Hilaire,
  Goriely, and Lechaftois}}]{Martini_2016}
\bibinfo{author}{\bibfnamefont{M.}~\bibnamefont{Martini}},
  \bibinfo{author}{\bibfnamefont{S.}~\bibnamefont{P{\'{e} }ru}},
  \bibinfo{author}{\bibfnamefont{S.}~\bibnamefont{Hilaire}},
  \bibinfo{author}{\bibfnamefont{S.}~\bibnamefont{Goriely}}, \bibnamefont{and}
  \bibinfo{author}{\bibfnamefont{F.}~\bibnamefont{Lechaftois}},
  \bibinfo{journal}{Physical Review C} \textbf{\bibinfo{volume}{94}}
  (\bibinfo{year}{2016}{\natexlab{a}}),
  \urlprefix\url{https://doi.org/10.1103%2Fphysrevc.94.014304}.

\bibitem[{\citenamefont{Kru\ifmmode \check{z}\else
  \v{z}\fi{}i\ifmmode~\acute{c}\else \'{c}\fi{}
  et~al.}(2021)\citenamefont{Kru\ifmmode \check{z}\else
  \v{z}\fi{}i\ifmmode~\acute{c}\else \'{c}\fi{}, Oishi, and Paar}}]{Krusiv2021}
\bibinfo{author}{\bibfnamefont{G.}~\bibnamefont{Kru\ifmmode \check{z}\else
  \v{z}\fi{}i\ifmmode~\acute{c}\else \'{c}\fi{}}},
  \bibinfo{author}{\bibfnamefont{T.}~\bibnamefont{Oishi}}, \bibnamefont{and}
  \bibinfo{author}{\bibfnamefont{N.}~\bibnamefont{Paar}},
  \bibinfo{journal}{Phys. Rev. C} \textbf{\bibinfo{volume}{103}},
  \bibinfo{pages}{054306} (\bibinfo{year}{2021}),
  \urlprefix\url{https://link.aps.org/doi/10.1103/PhysRevC.103.054306}.

\bibitem[{\citenamefont{Gambacurta et~al.}(2012)\citenamefont{Gambacurta,
  Grasso, De~Donno, Co', and Catara}}]{Gambacurta2012}
\bibinfo{author}{\bibfnamefont{D.}~\bibnamefont{Gambacurta}},
  \bibinfo{author}{\bibfnamefont{M.}~\bibnamefont{Grasso}},
  \bibinfo{author}{\bibfnamefont{V.}~\bibnamefont{De~Donno}},
  \bibinfo{author}{\bibfnamefont{G.}~\bibnamefont{Co'}}, \bibnamefont{and}
  \bibinfo{author}{\bibfnamefont{F.}~\bibnamefont{Catara}},
  \bibinfo{journal}{Phys. Rev. C} \textbf{\bibinfo{volume}{86}},
  \bibinfo{pages}{021304} (\bibinfo{year}{2012}),
  \urlprefix\url{https://link.aps.org/doi/10.1103/PhysRevC.86.021304}.

\bibitem[{\citenamefont{Gambacurta et~al.}(2015)\citenamefont{Gambacurta,
  Grasso, and Engel}}]{Gambacurta2015}
\bibinfo{author}{\bibfnamefont{D.}~\bibnamefont{Gambacurta}},
  \bibinfo{author}{\bibfnamefont{M.}~\bibnamefont{Grasso}}, \bibnamefont{and}
  \bibinfo{author}{\bibfnamefont{J.}~\bibnamefont{Engel}},
  \bibinfo{journal}{Phys. Rev. C} \textbf{\bibinfo{volume}{92}},
  \bibinfo{pages}{034303} (\bibinfo{year}{2015}),
  \urlprefix\url{https://link.aps.org/doi/10.1103/PhysRevC.92.034303}.

\bibitem[{\citenamefont{Knapp et~al.}(2023)\citenamefont{Knapp,
  Papakonstantinou, Vesel{\'{y} }, Gregorio, Herko, and Iudice}}]{Knapp_2023}
\bibinfo{author}{\bibfnamefont{F.}~\bibnamefont{Knapp}},
  \bibinfo{author}{\bibfnamefont{P.}~\bibnamefont{Papakonstantinou}},
  \bibinfo{author}{\bibfnamefont{P.}~\bibnamefont{Vesel{\'{y} }}},
  \bibinfo{author}{\bibfnamefont{G.~D.} \bibnamefont{Gregorio}},
  \bibinfo{author}{\bibfnamefont{J.}~\bibnamefont{Herko}}, \bibnamefont{and}
  \bibinfo{author}{\bibfnamefont{N.~L.} \bibnamefont{Iudice}},
  \bibinfo{journal}{Physical Review C} \textbf{\bibinfo{volume}{107}}
  (\bibinfo{year}{2023}),
  \urlprefix\url{https://doi.org/10.1103%2Fphysrevc.107.014305}.

\bibitem[{\citenamefont{Trippel}(2016)}]{trippel2016}
\bibinfo{author}{\bibfnamefont{R.}~\bibnamefont{Trippel}}, Ph.D. thesis,
  \bibinfo{school}{TU Darmstadt} (\bibinfo{year}{2016}), \bibinfo{note}{{D17,
  TU Darmstadt} (2016); https://tuprints.ulb.tudarmstadt.de/5883},
  \urlprefix\url{https://tuprints.ulb.tudarmstadt.de/5883}.

\bibitem[{\citenamefont{Goriely and Khan}(2002)}]{Goriely2002}
\bibinfo{author}{\bibfnamefont{S.}~\bibnamefont{Goriely}} \bibnamefont{and}
  \bibinfo{author}{\bibfnamefont{E.}~\bibnamefont{Khan}},
  \bibinfo{journal}{Nucl. Phys.} \textbf{\bibinfo{volume}{A706}},
  \bibinfo{pages}{217} (\bibinfo{year}{2002}).

\bibitem[{\citenamefont{Martini
  et~al.}(2016{\natexlab{b}})\citenamefont{Martini, P\'eru, Hilaire, Goriely,
  and Lechaftois}}]{Martini2016}
\bibinfo{author}{\bibfnamefont{M.}~\bibnamefont{Martini}},
  \bibinfo{author}{\bibfnamefont{S.}~\bibnamefont{P\'eru}},
  \bibinfo{author}{\bibfnamefont{S.}~\bibnamefont{Hilaire}},
  \bibinfo{author}{\bibfnamefont{S.}~\bibnamefont{Goriely}}, \bibnamefont{and}
  \bibinfo{author}{\bibfnamefont{F.}~\bibnamefont{Lechaftois}},
  \bibinfo{journal}{Phys. Rev. C} \textbf{\bibinfo{volume}{94}},
  \bibinfo{pages}{014304} (\bibinfo{year}{2016}{\natexlab{b}}).

\bibitem[{\citenamefont{Goriely et~al.}(2016)\citenamefont{Goriely, Hilaire,
  P\'eru, Martini, Deloncle, and Lechaftois}}]{Goriely2016}
\bibinfo{author}{\bibfnamefont{S.}~\bibnamefont{Goriely}},
  \bibinfo{author}{\bibfnamefont{S.}~\bibnamefont{Hilaire}},
  \bibinfo{author}{\bibfnamefont{S.}~\bibnamefont{P\'eru}},
  \bibinfo{author}{\bibfnamefont{M.}~\bibnamefont{Martini}},
  \bibinfo{author}{\bibfnamefont{I.}~\bibnamefont{Deloncle}}, \bibnamefont{and}
  \bibinfo{author}{\bibfnamefont{F.}~\bibnamefont{Lechaftois}},
  \bibinfo{journal}{Phys. Rev. C} \textbf{\bibinfo{volume}{94}},
  \bibinfo{pages}{044306} (\bibinfo{year}{2016}).

\bibitem[{\citenamefont{Goriely et~al.}(2018)\citenamefont{Goriely, Hilaire,
  P\'eru, and Sieja}}]{Goriely2018}
\bibinfo{author}{\bibfnamefont{S.}~\bibnamefont{Goriely}},
  \bibinfo{author}{\bibfnamefont{S.}~\bibnamefont{Hilaire}},
  \bibinfo{author}{\bibfnamefont{S.}~\bibnamefont{P\'eru}}, \bibnamefont{and}
  \bibinfo{author}{\bibfnamefont{K.}~\bibnamefont{Sieja}},
  \bibinfo{journal}{Phys. Rev. C} \textbf{\bibinfo{volume}{98}},
  \bibinfo{pages}{014327} (\bibinfo{year}{2018}).

\bibitem[{\citenamefont{Sieja and Goriely}(2021)}]{sieja2020directcapture}
\bibinfo{author}{\bibfnamefont{K.}~\bibnamefont{Sieja}} \bibnamefont{and}
  \bibinfo{author}{\bibfnamefont{S.}~\bibnamefont{Goriely}},
  \bibinfo{journal}{Eur. Phys. J. A} \textbf{\bibinfo{volume}{57}},
  \bibinfo{pages}{57} (\bibinfo{year}{2021}).

\bibitem[{\citenamefont{Stetcu and Johnson}(2003)}]{Stetcu2003}
\bibinfo{author}{\bibfnamefont{I.}~\bibnamefont{Stetcu}} \bibnamefont{and}
  \bibinfo{author}{\bibfnamefont{C.~W.} \bibnamefont{Johnson}},
  \bibinfo{journal}{Phys. Rev.} \textbf{\bibinfo{volume}{C67}},
  \bibinfo{pages}{044315} (\bibinfo{year}{2003}).

\bibitem[{\citenamefont{Caurier et~al.}(2005)\citenamefont{Caurier,
  Martinez-Pinedo, Nowacki, Poves, and Zuker}}]{RMP}
\bibinfo{author}{\bibfnamefont{E.}~\bibnamefont{Caurier}},
  \bibinfo{author}{\bibfnamefont{G.}~\bibnamefont{Martinez-Pinedo}},
  \bibinfo{author}{\bibfnamefont{F.}~\bibnamefont{Nowacki}},
  \bibinfo{author}{\bibfnamefont{A.}~\bibnamefont{Poves}}, \bibnamefont{and}
  \bibinfo{author}{\bibfnamefont{A.~P.} \bibnamefont{Zuker}},
  \bibinfo{journal}{Rev. Mod. Phys.} \textbf{\bibinfo{volume}{77}},
  \bibinfo{pages}{427} (\bibinfo{year}{2005}).

\bibitem[{\citenamefont{Loens et~al.}(2012)}]{Loens}
\bibinfo{author}{\bibfnamefont{H.}~\bibnamefont{Loens}} \bibnamefont{et~al.},
  \bibinfo{journal}{Eur. Phys. J.} \textbf{\bibinfo{volume}{48}},
  \bibinfo{pages}{34} (\bibinfo{year}{2012}).

\bibitem[{\citenamefont{Brown and Richter}(2006)}]{USDB}
\bibinfo{author}{\bibfnamefont{B.~A.} \bibnamefont{Brown}} \bibnamefont{and}
  \bibinfo{author}{\bibfnamefont{W.~A.} \bibnamefont{Richter}},
  \bibinfo{journal}{Phys. Rev. C} \textbf{\bibinfo{volume}{74}},
  \bibinfo{pages}{034315} (\bibinfo{year}{2006}).

\bibitem[{\citenamefont{Lenzi et~al.}(2010)\citenamefont{Lenzi, Nowacki, Poves,
  and Sieja}}]{Lenzi2010}
\bibinfo{author}{\bibfnamefont{S.~M.} \bibnamefont{Lenzi}},
  \bibinfo{author}{\bibfnamefont{F.}~\bibnamefont{Nowacki}},
  \bibinfo{author}{\bibfnamefont{A.}~\bibnamefont{Poves}}, \bibnamefont{and}
  \bibinfo{author}{\bibfnamefont{K.}~\bibnamefont{Sieja}},
  \bibinfo{journal}{Phys. Rev. C} \textbf{\bibinfo{volume}{82}},
  \bibinfo{pages}{054301} (\bibinfo{year}{2010}).

\bibitem[{\citenamefont{Gniady et~al.}()\citenamefont{Gniady, Caurier, Nowacki,
  and Poves}}]{Gniady}
\bibinfo{author}{\bibfnamefont{A.}~\bibnamefont{Gniady}},
  \bibinfo{author}{\bibfnamefont{E.}~\bibnamefont{Caurier}},
  \bibinfo{author}{\bibfnamefont{F.}~\bibnamefont{Nowacki}}, \bibnamefont{and}
  \bibinfo{author}{\bibfnamefont{A.}~\bibnamefont{Poves}},
  \bibinfo{howpublished}{unpublished}.

\bibitem[{\citenamefont{Caurier et~al.}(2010)\citenamefont{Caurier, Nowacki,
  Poves, and Sieja}}]{xenon}
\bibinfo{author}{\bibfnamefont{E.}~\bibnamefont{Caurier}},
  \bibinfo{author}{\bibfnamefont{F.}~\bibnamefont{Nowacki}},
  \bibinfo{author}{\bibfnamefont{A.}~\bibnamefont{Poves}}, \bibnamefont{and}
  \bibinfo{author}{\bibfnamefont{K.}~\bibnamefont{Sieja}},
  \bibinfo{journal}{Phys. Rev. C} \textbf{\bibinfo{volume}{82}},
  \bibinfo{pages}{064304} (\bibinfo{year}{2010}).

\bibitem[{\citenamefont{Sieja}(2018)}]{Sieja2018}
\bibinfo{author}{\bibfnamefont{K.}~\bibnamefont{Sieja}},
  \bibinfo{journal}{Phys. Rev. C} \textbf{\bibinfo{volume}{98}},
  \bibinfo{pages}{064312} (\bibinfo{year}{2018}).

\bibitem[{\citenamefont{Jakob et~al.}(2002)\citenamefont{Jakob, Benczer-Koller,
  Kumbartzki, Holden, Mertzimekis, Speidel, Ernst, Stuchbery, Pakou,
  Maier-Komor et~al.}}]{PhysRevC.65.024316}
\bibinfo{author}{\bibfnamefont{G.}~\bibnamefont{Jakob}},
  \bibinfo{author}{\bibfnamefont{N.}~\bibnamefont{Benczer-Koller}},
  \bibinfo{author}{\bibfnamefont{G.}~\bibnamefont{Kumbartzki}},
  \bibinfo{author}{\bibfnamefont{J.}~\bibnamefont{Holden}},
  \bibinfo{author}{\bibfnamefont{T.~J.} \bibnamefont{Mertzimekis}},
  \bibinfo{author}{\bibfnamefont{K.-H.} \bibnamefont{Speidel}},
  \bibinfo{author}{\bibfnamefont{R.}~\bibnamefont{Ernst}},
  \bibinfo{author}{\bibfnamefont{A.~E.} \bibnamefont{Stuchbery}},
  \bibinfo{author}{\bibfnamefont{A.}~\bibnamefont{Pakou}},
  \bibinfo{author}{\bibfnamefont{P.}~\bibnamefont{Maier-Komor}},
  \bibnamefont{et~al.}, \bibinfo{journal}{Phys. Rev. C}
  \textbf{\bibinfo{volume}{65}}, \bibinfo{pages}{024316}
  (\bibinfo{year}{2002}).

\bibitem[{\citenamefont{Caurier and Nowacki}(1999)}]{ANTOINE}
\bibinfo{author}{\bibfnamefont{E.}~\bibnamefont{Caurier}} \bibnamefont{and}
  \bibinfo{author}{\bibfnamefont{F.}~\bibnamefont{Nowacki}},
  \bibinfo{journal}{Acta Phys. Pol.} \textbf{\bibinfo{volume}{B30}},
  \bibinfo{pages}{705} (\bibinfo{year}{1999}).

\bibitem[{\citenamefont{Bartholomew et~al.}(1972)\citenamefont{Bartholomew,
  Earle, Ferguson, Knowles, and Lone}}]{Bartholomew}
\bibinfo{author}{\bibfnamefont{G.}~\bibnamefont{Bartholomew}},
  \bibinfo{author}{\bibfnamefont{E.}~\bibnamefont{Earle}},
  \bibinfo{author}{\bibfnamefont{A.}~\bibnamefont{Ferguson}},
  \bibinfo{author}{\bibfnamefont{J.}~\bibnamefont{Knowles}}, \bibnamefont{and}
  \bibinfo{author}{\bibfnamefont{M.}~\bibnamefont{Lone}},
  \bibinfo{journal}{Adv. Nucl. Phys.} \textbf{\bibinfo{volume}{7}},
  \bibinfo{pages}{229} (\bibinfo{year}{1972}).

\bibitem[{\citenamefont{Ryssens and Alhassid}(2021)}]{HF-SHELL}
\bibinfo{author}{\bibfnamefont{W.}~\bibnamefont{Ryssens}} \bibnamefont{and}
  \bibinfo{author}{\bibfnamefont{Y.}~\bibnamefont{Alhassid}},
  \bibinfo{journal}{Eur. Phys. J.} \textbf{\bibinfo{volume}{A 57}},
  \bibinfo{pages}{76} (\bibinfo{year}{2021}).

\bibitem[{\citenamefont{Sommermann}(1983)}]{SOM83}
\bibinfo{author}{\bibfnamefont{H.}~\bibnamefont{Sommermann}},
  \bibinfo{journal}{Annals of Physics} \textbf{\bibinfo{volume}{151}},
  \bibinfo{pages}{163} (\bibinfo{year}{1983}), ISSN \bibinfo{issn}{0003-4916},
  \urlprefix\url{https://www.sciencedirect.com/science/article/pii/0003491683903184}.

\bibitem[{\citenamefont{Ryssens and Alhassid}()}]{HF-SHELL2}
\bibinfo{author}{\bibfnamefont{W.}~\bibnamefont{Ryssens}} \bibnamefont{and}
  \bibinfo{author}{\bibfnamefont{Y.}~\bibnamefont{Alhassid}},
  \emph{\bibinfo{title}{Solution of the finite-temperature (q)rpa equations for
  shell-model hamiltonians: Hf-shell v2}}, \bibinfo{howpublished}{in
  preparation}.

\bibitem[{\citenamefont{Rowe}(1970)}]{Rowe70}
\bibinfo{author}{\bibfnamefont{D.}~\bibnamefont{Rowe}},
  \emph{\bibinfo{title}{Nuclear collective motion}}
  (\bibinfo{publisher}{Methuen and Co., London}, \bibinfo{year}{1970}).

\bibitem[{\citenamefont{Nakatsukasa et~al.}(2007)\citenamefont{Nakatsukasa,
  Inakura, and Yabana}}]{Nakatsukasa2007}
\bibinfo{author}{\bibfnamefont{T.}~\bibnamefont{Nakatsukasa}},
  \bibinfo{author}{\bibfnamefont{T.}~\bibnamefont{Inakura}}, \bibnamefont{and}
  \bibinfo{author}{\bibfnamefont{K.}~\bibnamefont{Yabana}},
  \bibinfo{journal}{Phys. Rev. C} \textbf{\bibinfo{volume}{76}},
  \bibinfo{pages}{024318} (\bibinfo{year}{2007}),
  \urlprefix\url{https://link.aps.org/doi/10.1103/PhysRevC.76.024318}.

\bibitem[{\citenamefont{Avogadro and Nakatsukasa}(2011)}]{Nakatsukasa2011}
\bibinfo{author}{\bibfnamefont{P.}~\bibnamefont{Avogadro}} \bibnamefont{and}
  \bibinfo{author}{\bibfnamefont{T.}~\bibnamefont{Nakatsukasa}},
  \bibinfo{journal}{Phys. Rev. C} \textbf{\bibinfo{volume}{84}},
  \bibinfo{pages}{014314} (\bibinfo{year}{2011}),
  \urlprefix\url{https://link.aps.org/doi/10.1103/PhysRevC.84.014314}.

\bibitem[{\citenamefont{Inakura et~al.}(2009)\citenamefont{Inakura,
  Nakatsukasa, and Yabana}}]{Inakura2009}
\bibinfo{author}{\bibfnamefont{T.}~\bibnamefont{Inakura}},
  \bibinfo{author}{\bibfnamefont{T.}~\bibnamefont{Nakatsukasa}},
  \bibnamefont{and} \bibinfo{author}{\bibfnamefont{K.}~\bibnamefont{Yabana}},
  \bibinfo{journal}{Phys. Rev. C} \textbf{\bibinfo{volume}{80}},
  \bibinfo{pages}{044301} (\bibinfo{year}{2009}),
  \urlprefix\url{https://link.aps.org/doi/10.1103/PhysRevC.80.044301}.

\bibitem[{\citenamefont{Stoitsov et~al.}(2011)\citenamefont{Stoitsov,
  Kortelainen, Nakatsukasa, Losa, and Nazarewicz}}]{Stoitsov2011}
\bibinfo{author}{\bibfnamefont{M.}~\bibnamefont{Stoitsov}},
  \bibinfo{author}{\bibfnamefont{M.}~\bibnamefont{Kortelainen}},
  \bibinfo{author}{\bibfnamefont{T.}~\bibnamefont{Nakatsukasa}},
  \bibinfo{author}{\bibfnamefont{C.}~\bibnamefont{Losa}}, \bibnamefont{and}
  \bibinfo{author}{\bibfnamefont{W.}~\bibnamefont{Nazarewicz}},
  \bibinfo{journal}{Phys. Rev. C} \textbf{\bibinfo{volume}{84}},
  \bibinfo{pages}{041305} (\bibinfo{year}{2011}),
  \urlprefix\url{https://link.aps.org/doi/10.1103/PhysRevC.84.041305}.

\bibitem[{\citenamefont{Ney et~al.}(2020)\citenamefont{Ney, Engel, Li, and
  Schunck}}]{Ney2020}
\bibinfo{author}{\bibfnamefont{E.~M.} \bibnamefont{Ney}},
  \bibinfo{author}{\bibfnamefont{J.}~\bibnamefont{Engel}},
  \bibinfo{author}{\bibfnamefont{T.}~\bibnamefont{Li}}, \bibnamefont{and}
  \bibinfo{author}{\bibfnamefont{N.}~\bibnamefont{Schunck}},
  \bibinfo{journal}{Phys. Rev. C} \textbf{\bibinfo{volume}{102}},
  \bibinfo{pages}{034326} (\bibinfo{year}{2020}),
  \urlprefix\url{https://link.aps.org/doi/10.1103/PhysRevC.102.034326}.

\bibitem[{\citenamefont{Mustonen et~al.}(2014)\citenamefont{Mustonen, Shafer,
  Zenginerler, and Engel}}]{Mustonen2014}
\bibinfo{author}{\bibfnamefont{M.~T.} \bibnamefont{Mustonen}},
  \bibinfo{author}{\bibfnamefont{T.}~\bibnamefont{Shafer}},
  \bibinfo{author}{\bibfnamefont{Z.}~\bibnamefont{Zenginerler}},
  \bibnamefont{and} \bibinfo{author}{\bibfnamefont{J.}~\bibnamefont{Engel}},
  \bibinfo{journal}{Phys. Rev. C} \textbf{\bibinfo{volume}{90}},
  \bibinfo{pages}{024308} (\bibinfo{year}{2014}),
  \urlprefix\url{https://link.aps.org/doi/10.1103/PhysRevC.90.024308}.

\bibitem[{\citenamefont{Washiyama et~al.}(2021)\citenamefont{Washiyama,
  Hinohara, and Nakatsukasa}}]{Washiyama2021}
\bibinfo{author}{\bibfnamefont{K.}~\bibnamefont{Washiyama}},
  \bibinfo{author}{\bibfnamefont{N.}~\bibnamefont{Hinohara}}, \bibnamefont{and}
  \bibinfo{author}{\bibfnamefont{T.}~\bibnamefont{Nakatsukasa}},
  \bibinfo{journal}{Phys. Rev. C} \textbf{\bibinfo{volume}{103}},
  \bibinfo{pages}{014306} (\bibinfo{year}{2021}),
  \urlprefix\url{https://link.aps.org/doi/10.1103/PhysRevC.103.014306}.

\bibitem[{\citenamefont{Litvinova and Zhang}(2021)}]{Litvinova2021}
\bibinfo{author}{\bibfnamefont{E.}~\bibnamefont{Litvinova}} \bibnamefont{and}
  \bibinfo{author}{\bibfnamefont{Y.}~\bibnamefont{Zhang}},
  \bibinfo{journal}{Phys. Rev. C} \textbf{\bibinfo{volume}{104}},
  \bibinfo{pages}{044303} (\bibinfo{year}{2021}),
  \urlprefix\url{https://link.aps.org/doi/10.1103/PhysRevC.104.044303}.

\bibitem[{\citenamefont{Hinohara et~al.}(2013)\citenamefont{Hinohara,
  Kortelainen, and Nazarewicz}}]{Hinohara2013}
\bibinfo{author}{\bibfnamefont{N.}~\bibnamefont{Hinohara}},
  \bibinfo{author}{\bibfnamefont{M.}~\bibnamefont{Kortelainen}},
  \bibnamefont{and}
  \bibinfo{author}{\bibfnamefont{W.}~\bibnamefont{Nazarewicz}},
  \bibinfo{journal}{Phys. Rev. C} \textbf{\bibinfo{volume}{87}},
  \bibinfo{pages}{064309} (\bibinfo{year}{2013}),
  \urlprefix\url{https://link.aps.org/doi/10.1103/PhysRevC.87.064309}.

\bibitem[{\citenamefont{Hinohara}(2015)}]{Hinohara2015}
\bibinfo{author}{\bibfnamefont{N.}~\bibnamefont{Hinohara}},
  \bibinfo{journal}{Phys. Rev. C} \textbf{\bibinfo{volume}{92}},
  \bibinfo{pages}{034321} (\bibinfo{year}{2015}),
  \urlprefix\url{https://link.aps.org/doi/10.1103/PhysRevC.92.034321}.

\bibitem[{\citenamefont{Ring and Schuck}(1980{\natexlab{b}})}]{RingSchuck}
\bibinfo{author}{\bibfnamefont{P.}~\bibnamefont{Ring}} \bibnamefont{and}
  \bibinfo{author}{\bibfnamefont{P.}~\bibnamefont{Schuck}},
  \emph{\bibinfo{title}{The Nuclear Many-Body Problems}}, vol.
  \bibinfo{volume}{103} (\bibinfo{year}{1980}{\natexlab{b}}).

\bibitem[{\citenamefont{Duguet et~al.}(2020)\citenamefont{Duguet, Bally, and
  Tichai}}]{duguet2020}
\bibinfo{author}{\bibfnamefont{T.}~\bibnamefont{Duguet}},
  \bibinfo{author}{\bibfnamefont{B.}~\bibnamefont{Bally}}, \bibnamefont{and}
  \bibinfo{author}{\bibfnamefont{A.}~\bibnamefont{Tichai}},
  \bibinfo{journal}{Phys. Rev. C} \textbf{\bibinfo{volume}{102}},
  \bibinfo{pages}{054320} (\bibinfo{year}{2020}), ISSN
  \bibinfo{issn}{2469-9985, 2469-9993}, \eprint{2006.02871}.

\bibitem[{\citenamefont{Duguet and Ryssens}(2020)}]{duguet2020a}
\bibinfo{author}{\bibfnamefont{T.}~\bibnamefont{Duguet}} \bibnamefont{and}
  \bibinfo{author}{\bibfnamefont{W.}~\bibnamefont{Ryssens}},
  \bibinfo{journal}{Phys. Rev. C} \textbf{\bibinfo{volume}{102}},
  \bibinfo{pages}{044328} (\bibinfo{year}{2020}).

\bibitem[{\citenamefont{Porro}(In preparation)}]{porrothesis}
\bibinfo{author}{\bibfnamefont{A.}~\bibnamefont{Porro}} (\bibinfo{year}{In
  preparation}).

\bibitem[{\citenamefont{Federschmidt and Ring}(1985)}]{Federschmidt1985}
\bibinfo{author}{\bibfnamefont{C.}~\bibnamefont{Federschmidt}}
  \bibnamefont{and} \bibinfo{author}{\bibfnamefont{P.}~\bibnamefont{Ring}},
  \bibinfo{journal}{Nuclear Physics A} \textbf{\bibinfo{volume}{435}},
  \bibinfo{pages}{110} (\bibinfo{year}{1985}), ISSN \bibinfo{issn}{0375-9474},
  \urlprefix\url{https://www.sciencedirect.com/science/article/pii/0375947485903070}.

\bibitem[{\citenamefont{Lipparini}(2003)}]{lipparini2003}
\bibinfo{author}{\bibfnamefont{E.}~\bibnamefont{Lipparini}}, in
  \emph{\bibinfo{booktitle}{Modern {{Many-Particle Physics}}: {{Atomic Gases}},
  {{Quantum Dots}} and {{Quantum Fluids}} by {{Enrico Lipparini}}
  ({{Author}})}} (\bibinfo{year}{2003}).

\bibitem[{\citenamefont{Gaudefroy et~al.}(2018)\citenamefont{Gaudefroy, Péru,
  Arnal, Aupiais, Delaroche, Girod, and Libert}}]{Gaudefroy2018}
\bibinfo{author}{\bibfnamefont{L.}~\bibnamefont{Gaudefroy}},
  \bibinfo{author}{\bibfnamefont{S.}~\bibnamefont{Péru}},
  \bibinfo{author}{\bibfnamefont{N.}~\bibnamefont{Arnal}},
  \bibinfo{author}{\bibfnamefont{J.}~\bibnamefont{Aupiais}},
  \bibinfo{author}{\bibfnamefont{J.-P.} \bibnamefont{Delaroche}},
  \bibinfo{author}{\bibfnamefont{M.}~\bibnamefont{Girod}}, \bibnamefont{and}
  \bibinfo{author}{\bibfnamefont{J.}~\bibnamefont{Libert}},
  \bibinfo{journal}{Phys. Rev. C} \textbf{\bibinfo{volume}{97}},
  \bibinfo{pages}{064317} (\bibinfo{year}{2018}), \bibinfo{note}{publisher:
  American Physical Society},
  \urlprefix\url{https://link.aps.org/doi/10.1103/PhysRevC.97.064317}.

\bibitem[{\citenamefont{Tsoneva and Lenske}(2008)}]{tsoneva2008}
\bibinfo{author}{\bibfnamefont{N.}~\bibnamefont{Tsoneva}} \bibnamefont{and}
  \bibinfo{author}{\bibfnamefont{H.}~\bibnamefont{Lenske}},
  \bibinfo{journal}{Phys. Rev. C} \textbf{\bibinfo{volume}{77}},
  \bibinfo{pages}{024321} (\bibinfo{year}{2008}).

\bibitem[{\citenamefont{Rusev et~al.}(2013)\citenamefont{Rusev, Tsoneva,
  D\"onau, Frauendorf, Schwengner, Tonchev, Adekola, Hammond, Kelley, Kwan
  et~al.}}]{Rusev2013}
\bibinfo{author}{\bibfnamefont{G.}~\bibnamefont{Rusev}},
  \bibinfo{author}{\bibfnamefont{N.}~\bibnamefont{Tsoneva}},
  \bibinfo{author}{\bibfnamefont{F.}~\bibnamefont{D\"onau}},
  \bibinfo{author}{\bibfnamefont{S.}~\bibnamefont{Frauendorf}},
  \bibinfo{author}{\bibfnamefont{R.}~\bibnamefont{Schwengner}},
  \bibinfo{author}{\bibfnamefont{A.~P.} \bibnamefont{Tonchev}},
  \bibinfo{author}{\bibfnamefont{A.~S.} \bibnamefont{Adekola}},
  \bibinfo{author}{\bibfnamefont{S.~L.} \bibnamefont{Hammond}},
  \bibinfo{author}{\bibfnamefont{J.~H.} \bibnamefont{Kelley}},
  \bibinfo{author}{\bibfnamefont{E.}~\bibnamefont{Kwan}}, \bibnamefont{et~al.},
  \bibinfo{journal}{Phys. Rev. Lett.} \textbf{\bibinfo{volume}{110}},
  \bibinfo{pages}{022503} (\bibinfo{year}{2013}),
  \urlprefix\url{https://link.aps.org/doi/10.1103/PhysRevLett.110.022503}.

\bibitem[{\citenamefont{Duguet et~al.}(2022)\citenamefont{Duguet, Ebran,
  Frosini, Hergert, and Som{\`a}}}]{duguet2022}
\bibinfo{author}{\bibfnamefont{T.}~\bibnamefont{Duguet}},
  \bibinfo{author}{\bibfnamefont{J.-P.} \bibnamefont{Ebran}},
  \bibinfo{author}{\bibfnamefont{M.}~\bibnamefont{Frosini}},
  \bibinfo{author}{\bibfnamefont{H.}~\bibnamefont{Hergert}}, \bibnamefont{and}
  \bibinfo{author}{\bibfnamefont{V.}~\bibnamefont{Som{\`a}}},
  \emph{\bibinfo{title}{Rooting the {{EDF}} method into the ab initio
  framework. {{PGCM-PT}} formalism based on {{MR-IMSRG}} pre-processed
  {{Hamiltonians}}}} (\bibinfo{year}{2022}), \eprint{2209.03424}.

\bibitem[{\citenamefont{Koonin et~al.}(1997)\citenamefont{Koonin, Dean, and
  Langanke}}]{Koonin1997}
\bibinfo{author}{\bibfnamefont{S.~E.} \bibnamefont{Koonin}},
  \bibinfo{author}{\bibfnamefont{D.~J.} \bibnamefont{Dean}}, \bibnamefont{and}
  \bibinfo{author}{\bibfnamefont{K.}~\bibnamefont{Langanke}},
  \bibinfo{journal}{Physics Reports} \textbf{\bibinfo{volume}{278}},
  \bibinfo{pages}{1} (\bibinfo{year}{1997}), ISSN \bibinfo{issn}{0370-1573}.

\bibitem[{\citenamefont{Fanto and Alhassid}(2021)}]{fanto2021}
\bibinfo{author}{\bibfnamefont{P.}~\bibnamefont{Fanto}} \bibnamefont{and}
  \bibinfo{author}{\bibfnamefont{Y.}~\bibnamefont{Alhassid}},
  \emph{\bibinfo{title}{Low-energy enhancement in the magnetic dipole
  \${\textbackslash}gamma\$-ray strength functions of heavy nuclei}}
  (\bibinfo{year}{2021}), \bibinfo{note}{arXiv:2112.13772 [cond-mat,
  physics:nucl-th]}, \urlprefix\url{http://arxiv.org/abs/2112.13772}.

\bibitem[{\citenamefont{Shimizu et~al.}(2015)\citenamefont{Shimizu, Abe, Honma,
  Otsuka, Tsunoda, Utsuno, and Yoshida}}]{Shimizu-Ca48}
\bibinfo{author}{\bibfnamefont{N.}~\bibnamefont{Shimizu}},
  \bibinfo{author}{\bibfnamefont{T.}~\bibnamefont{Abe}},
  \bibinfo{author}{\bibfnamefont{M.}~\bibnamefont{Honma}},
  \bibinfo{author}{\bibfnamefont{T.}~\bibnamefont{Otsuka}},
  \bibinfo{author}{\bibfnamefont{Y.}~\bibnamefont{Tsunoda}},
  \bibinfo{author}{\bibfnamefont{Y.}~\bibnamefont{Utsuno}}, \bibnamefont{and}
  \bibinfo{author}{\bibfnamefont{T.}~\bibnamefont{Yoshida}},
  \bibinfo{journal}{JPS Conf. Proc.} \textbf{\bibinfo{volume}{6}},
  \bibinfo{pages}{010021} (\bibinfo{year}{2015}).

\bibitem[{\citenamefont{Tsoneva et~al.}(2004)\citenamefont{Tsoneva, Lenske, and
  Stoyanov}}]{Tsoneva2004}
\bibinfo{author}{\bibfnamefont{N.}~\bibnamefont{Tsoneva}},
  \bibinfo{author}{\bibfnamefont{H.}~\bibnamefont{Lenske}}, \bibnamefont{and}
  \bibinfo{author}{\bibfnamefont{C.}~\bibnamefont{Stoyanov}},
  \bibinfo{journal}{Physics Letters B} \textbf{\bibinfo{volume}{586}},
  \bibinfo{pages}{213} (\bibinfo{year}{2004}), ISSN \bibinfo{issn}{0370-2693},
  \urlprefix\url{https://www.sciencedirect.com/science/article/pii/S0370269304003090}.

\bibitem[{\citenamefont{Bofos et~al.}()\citenamefont{Bofos, Martinez-Larraz,
  Bally, Duguet, Frosini, Rodriguez, and Sieja}}]{Bofos}
\bibinfo{author}{\bibfnamefont{S.}~\bibnamefont{Bofos}},
  \bibinfo{author}{\bibfnamefont{J.}~\bibnamefont{Martinez-Larraz}},
  \bibinfo{author}{\bibfnamefont{B.}~\bibnamefont{Bally}},
  \bibinfo{author}{\bibfnamefont{T.}~\bibnamefont{Duguet}},
  \bibinfo{author}{\bibfnamefont{M.}~\bibnamefont{Frosini}},
  \bibinfo{author}{\bibfnamefont{T.}~\bibnamefont{Rodriguez}},
  \bibnamefont{and} \bibinfo{author}{\bibfnamefont{K.}~\bibnamefont{Sieja}},
  \emph{\bibinfo{title}{Application of projected generator coordinate method
  for m1 strength function calculation in valence space}},
  \bibinfo{howpublished}{in preparation}.

\end{thebibliography}
\end{document}